\documentclass[twocolumn,aps,prb,superscriptaddress,showkeys,showpacs,10pt]{revtex4-1}
\usepackage{graphicx}% Include figure files
\usepackage{dcolumn}% Align table columns on decimal point
\usepackage{bm}% bold math[;[
\usepackage{subfigure}
\usepackage{multirow}
\usepackage{gensymb}
\usepackage[usenames,dvipsnames]{color}
\usepackage{soul}
\usepackage[colorlinks=true,allcolors=blue]{hyperref}
\usepackage[all]{hypcap}    %for going to the top of an image when a figure reference is clicked

\usepackage{amsmath}
\usepackage{xfrac}
\usepackage{nicefrac}
\usepackage{ulem}

\newcommand\BHO{BaHo$_2$O$_4$}
\newcommand\BDO{BaDy$_2$O$_4$}
\newcommand\SHO{SrHo$_2$O$_4$}
\newcommand\SDO{SrDy$_2$O$_4$}
\newcommand\SLnO{Sr\textit{Ln}$_2$O$_4$}
\newcommand\SDHO{Sr(Dy,Ho)$_2$O$_4$}

\newcommand\Tstrut{\rule{0pt}{2.6ex}}         % = `top' strut
\newcommand\Bstrut{\rule[-1.2ex]{0pt}{0pt}}

\begin{document}

%%%%%%%%%%%%%%%%%%%%%%%%%%%%%%%%%%%%%%%%%%%%%%%%%%%%%%%%%%%%%%%%%%%%%%%%%%%%%%%%%%%%%%%%%%%%%%%%%%%%%%%%%%%%%%%%%%%%%%%%%%%%%%%%%%%%%%%%%%%%%%%%%%%
\title{Coexistence of magnetic fluctuations and long-range orders in the one-dimensional \texorpdfstring{$J_1$}{}--\texorpdfstring{$J_2$}{} zigzag chains materials \texorpdfstring{\BDO}{} and \texorpdfstring{\BHO}{}} %%%%%
%%%%%%%%%%%%%%%%%%%%%%%%%%%%%%%%%%%%%%%%%%%%%%%%%%%%%%%%%%%%%%%%%%%%%%%%%%%%%%%%%%%%%%%%%%%%%%%%%%%%%%%%%%%%%%%%%%%%%%%%%%%%%%%%%%%%%%%%%%%%%%%%%%%

\author{Bobby Pr\'evost}
\affiliation{D\'epartement de Physique \& Regroupement Qu\'eb\'ecois sur les Mat\'eriaux de Pointe (RQMP), Universit\'e de Montr\'eal, Montr\'eal, Qu\'ebec, Canada H3C 3J7}

\author{Nicolas Gauthier}
\affiliation{Laboratory for Scientific Developments and Novel Materials, Paul Scherrer Institut, CH-5232 Villigen PSI, Switzerland.}
 \altaffiliation[Present address: ]{Stanford Institute for Materials and Energy Sciences, SLAC National Accelerator Laboratory, Menlo Park, California 94025, USA.}

\author{Vladimir Y. Pomjakushin}
\affiliation{Laboratory for Neutron Scattering, Paul Scherrer Institute, CH-5232 Villigen, Switzerland}

\author{Bernard Delley}
\affiliation{Condensed Matter Theory Group, Paul Scherrer Institut, CH-5232 Villigen PSI, Switzerland}

\author{Helen C. Walker}
\affiliation{ISIS Facility, STFC Rutherford Appleton Laboratory, Chilton, Didcot, Oxfordshire OX11 0QX, United Kingdom}

\author{Michel Kenzelmann}
\affiliation{Laboratory for Neutron Scattering, Paul Scherrer Institute, CH-5232 Villigen, Switzerland}

\author{Andrea D. Bianchi}
\email[Corresponding address: ]{andrea.bianchi@umontreal.ca}
\affiliation{D\'epartement de Physique \& Regroupement Qu\'eb\'ecois sur les Mat\'eriaux de Pointe (RQMP), Universit\'e de Montr\'eal, Montr\'eal, Qu\'ebec, Canada H3C 3J7}

\date{\today}

% \keywords{Magnetic frustration, short-range order, low-dimensionality, neutron scattering, crystal field}

%%%%%%%%%%%%%%%%%%%%%%
\begin{abstract} %%%%%
%%%%%%%%%%%%%%%%%%%%%%
The compounds \BDO{} and \BHO{} are part of a family of frustrated systems exhibiting interesting properties, including spin liquid-type ground states, magnetic field-induced phases, and the coexistence of short- and long-range magnetic orders, with dominant one-dimensional correlations, which can be described as Ising $J_1$--$J_2$ zigzag chains along the $c$-axis. We have investigated polycrystalline samples of \BDO{} and \BHO{} with both neutron diffraction and neutron spectroscopy, coupled to detailed crystalline electric field calculations. The latter points to site-dependent anisotropic magnetism in both materials, which is corroborated by the magnetic structures we determined. The two systems show the coexistence of two different long-range orders --- two double N\'eel $\uparrow\uparrow\downarrow\downarrow$ orders in the $ab$-plane with propagation vectors $\mathbf{k}_1$~=~($\frac{1}{2}$,0,$\frac{1}{2}$) and $\mathbf{k}_2$~=~($\frac{1}{2}$,$\frac{1}{2}$,$\frac{1}{2}$) for \BDO{}, and two distinct arrangements of simple N\'eel $\uparrow\downarrow\uparrow\downarrow$ orders along the $c$-axis, both with the propagation vector $\mathbf{k}_0$~=~(0,0,0) for \BHO{}. The order for both wave vectors in \BDO{} occurs at $T_\mathrm{N}$~=~0.48~K, while in \BHO{}, the first order sets in at $T_\mathrm{N}\sim$~1.3~K and the second one has a lower ordering temperature of 0.84~K. Both compounds show extensive diffuse scattering which we successfully modeled with a one-dimensional axial next-nearest neighbor Ising (ANNNI) model. In both materials, strong diffusive scattering persists to temperatures well below where the magnetic order is fully saturated. The ANNNI model fits indicate the presence of sites which do not order with moments in the $ab$-plane. 

\end{abstract}

\maketitle

%%%%%%%%%%%%%%%%%%%%%%%%%%%%
\section{Introduction} %%%%%
%%%%%%%%%%%%%%%%%%%%%%%%%%%%

Among the pyrochlores, spinels, and kagome structures, a newly synthesized family of magnetically frustrated compounds with general formula \textit{AkLn}$_2$O$_4$ (where \textit{Ak}~=~alkaline earth metal, and {\textit{Ln}}~=~lanthanide) has recently attracted attention \cite{Karunadasa2005p41}, as they offer a novel route to magnetic frustration. This family of compounds shows a large variety of magnetic properties, ranging from the coexistence of short- and long-range magnetic order \cite{Petrenko2008p612,Ghosh:2011ew,Hayes:2011gf,Young:2012ji,Young:2013hw,Wen:2015wl}, to the complete absence of magnetic order down to the lowest accessible temperature despite strong magnetic interactions \cite{Fennell:2014fy,Li:2015dh,Gauthier:2017vz}. In applied fields, the magnetization shows the formation of plateaus at 1/3 of the saturation value \cite{Hayes:2012kt}, which are concurrent with field induced magnetic order \cite{QuinteroCastro:2012dd,Cheffings:2013bw,Bidaud:2016ki,Petrenko:2017em,Gauthier:2017ts}.

The crystalline structure of the \textit{AkLn}$_2$O$_4$ compounds belongs to the {\textit{Pnam}} space group, where the lattice of magnetic rare earth ions forms hexagonal tiles in the $ab$-plane, and zigzag chains along the $c$-axis. In this structure, each rare earth ion is located inside a differently distorted O octahedron, resulting in two inequivalent rare earth sites with a distinct crystal electric field (CEF) environment. It has been argued that the unusual magnetism in \SLnO{} originates from the chains, which can be described as an effective one-dimensional (1D) Ising $J_1$--$J_2$ spin chain model \cite{Fennell:2014gf,Wen:2015wl} --- commonly referred as the 1D axial next-nearest neighbor Ising (ANNNI) model \cite{Selke:1988tx}. The spin anisotropy in \textit{AkLn}$_2$O$_4$ is controlled by the CEF, which splits the $J$ multiplet of the rare earth ions \cite{Fennell:2014fy}. The observed CEF splitting often results in an Ising-type interaction between the rare earth moments, reducing the effective dimension of the exchange to a zigzag chain of spins.

Microscopic models trying to explain ferromagnetic order started out in one dimension, when Ising tried to determine the origin of magnetic order in ferromagnets \cite{Schollwock:2004wr,Ising:1925eg}. This development was followed by Heisenberg's discovery that the origin of the molecular fields in the Weiss theory of ferromagnetism are due to a combination of Coulomb repulsion and the Pauli exclusion principle. At the same time this also gave rise to the possibility of antiferromagnetic interactions, as well as the formation of singlets with spin zero starting from two interacting spin-1/2 \cite{Heisenberg:1928jw}. The quantum mechanical solution of the 1D antiferromagnetic Heisenberg model was later found by Bethe \cite{Bethe:1931iz}. After this, it took almost 40 years before the remarkable lack of long-range order at finite temperatures for systems with continuous symmetry in one and two dimensions was put on a solid theoretical footing by Mermin and Wagner \cite{Mermin:1966da}.

The discovery that the excitations of the 1D spin-1/2 Heisenberg chain are fermions by Faddeev and Takhtajan \cite{Faddeev:1981cl}, the so-called spinons, led to renewed interest in this field. This work was followed by the seminal papers of Haldane which showed the difference in the excitation spectrum between integer and half-integer 1D chains \cite{Haldane:1983cl,Haldane:1983ip}. The attraction to study low-dimensional spin systems became stronger after the discovery of the high-temperature superconductors which started the search for other such systems in the hope to increase the magnetic fluctuations, and as consequence, the superconducting critical temperature $T_c$. The result of these efforts, for example, were the so-called ladder compounds by Dagotto and Rice \cite{Dagotto:1996jl}, which are not only in between one and two dimensions, but also pointed to the important role frustration has in selecting the ground state. Then, Affleck discovered that the application of a magnetic field to integer spin chains in one dimension with antiferromagnetic interactions leads to a quantum phase transition where spin excitations undergo a Bose-Einstein condensation, initiating the search for new quantum magnetic ground states \cite{Affleck:1990cw}. Furthermore, the magnetization in these systems shows plateaus, as the magnetization per site is topologically quantized \cite{Oshikawa:1997fx}.

Antiferromagnetic zigzag chains, such as \textit{AkLn}$_2$O$_4$, combine the effect of low dimensionality with magnetic frustration due to the antiferromagnetic next-nearest neighbor interactions. In an applied magnetic fields, zigzag chains show a plateau in the magnetization at 1/3 of the saturation value, even in the classical limit \cite{HeidrichMeisner:2007ey}, and a rich magnetic phase diagram \cite{Schollwock:2004wr}. Also, while conventional long-range order is prohibited by the Mermin-Wagner theorem, zigzag chains can still acquire chiral order, which should survive to finite temperatures \cite{McCulloch:2008cw}. In a chirally ordered state, spins have a tendency to “rotate” in a preferred plane with a preferred rotational direction \cite{McCulloch:2008cw}.

In the ANNNI model, the ground state at zero temperature is determined through competing interactions between the nearest neighbors $J_1$ and the next-nearest neighbors $J_2$:
\begin{equation}
{\cal H} = \sum_i J_1{\hat{S}_i^z}{\hat{S}_{i+1}^z} + J_2{\hat{S}_i^z}{\hat{S}_{i+2}^z}. 
\end{equation}
The ANNNI model can accommodate different types of anti-ferromagnetic order, depending on the ratio $J_2/J_1$ of the interactions \cite{Okunishi:2003ux}. The ground state is a simple N\'eel state for an antiferromagnetic $J_1$ and a $J_2$ which is either ferromagnetic or weakly anti-ferromagnetic, such that $J_2 > -0.5|J_1|$. A double N\'eel state forms for an antiferromagnetic $J_2$ with $J_2 < -0.5|J_1|$, where $J_1$ can be either ferromagnetic, or antiferromagnetic. A ratio close to the critical value separating these two ground states leads to frustration --- a strong source for fluctuations.

In this paper, we present a detailed investigation of the magnetic structure of \BDO{} and \BHO{} for which only growth, specific heat, and magnetization have been reported \cite{Doi:2006ue,Besara:2014kx}. The motivation behind this investigation is a direct comparison with the isotructural compounds \SDHO{} by replacing the non-interacting Sr atoms with larger Ba atoms. As the interactions are governed by crystalline electric field effects, which are very sensitive to local structure, the change from Sr to Ba will change the anisotropy and strength of the magnetic interactions.

%%%%%%%%%%%%%%%%%%%%%%%%%%%%%%%%%%%%
\section{Results and analysis} %%%%%
%%%%%%%%%%%%%%%%%%%%%%%%%%%%%%%%%%%%

%%%%%%%%%%%%%%%%%%%%%%%%%%%%%%%%%%%%%%%%%%%%%%%%%%%%%%%%%%%%%%%%%
\subsection{Samples preparation and experimental technique} %%%%%
%%%%%%%%%%%%%%%%%%%%%%%%%%%%%%%%%%%%%%%%%%%%%%%%%%%%%%%%%%%%%%%%%

\BDO{} and \BHO{} powders were prepared by solid state reaction using high purity starting materials. Stoichiometric mixtures of Dy$_2$O$_3$/Ho$_2$O$_3$ (99.995\%) and BaCO$_3$ (99.994\%), with a 1\% surplus of carbonate, both dried at 600$\degree$C and weighed in a glove-box, were mixed in a ball mill and pressed into a rod. The rods were heated in an argon atmosphere at 1300$\degree$C in an alumina crucible for 12 hours. The solidified rods were ground, repressed into rods, and heated once more under the same condition. This resulted in single phase samples, as determined by X-ray powder diffraction. For specific heat measurements, small single crystals of \BHO{} were grown from Ba flux following the method described by Besara {\textit{et al.}}~\cite{Besara:2014kx}.

In order to establish the CEF splitting of the ground state multiplets $J$~=~15/2 and $J$~=~8 of the Dy$^{3+}$ and the Ho$^{3+}$ ions, respectively, inelastic neutron scattering experiments were performed on powders of both \BDO{} and \BHO{} using the MERLIN time-of-flight spectrometer at ISIS in the United-Kingdom \cite{Bewley:2006bl}. The two samples were mounted in a double-walled aluminum cylinder to maximize the scattering intensities \cite{Schmitt:1998by}. Spectra were acquired for both samples at incident neutron energies $E_i$ of 12.30, 20.00, 38.20 and 100.00~meV and temperatures $T$ of 7, 75 and 150~K. A calculation of the CEF energy levels and the inelastic neutron spectra were performed with the program \textsc{multiX} \cite{Uldry:2012ky}.

To determine the magnetic structure of the compounds, elastic neutron scattering data were acquired on the same powders using the HRPT diffractometer at the Paul Scherrer Institut (PSI) in Switzerland \cite{Fischer:2000ja}. Data were collected from $T$~=~0.07~K up to 30~K at a neutron wavelength $\lambda$ of 1.886~\AA{} for \BDO{}, and from $T$~=~0.20~K up to 60~K at the same wavelength, with additional data at $\lambda$~=~1.155~\AA{}, for \BHO{}. Due to the strong neutron absorption of Dy, the \BDO{} powder was loaded into a double walled copper cylinder, while for \BHO{}, a simple copper cylinder was used. Both cylinders were filled with 10~bar of helium at ambient temperature to ensure a good thermalization. The magnetic structure was refined through a Rietveld analysis with the \textsc{FullProf} Suite \cite{RodriguezCarvajal:1993cf}.

Magnetization measurements on the powders were carried out using a Quantum Design SQUID-VSM, equipped with a 7~T magnet, over the temperature range from 300~K down to 1.8~K. The specific heat measurements were made using a Quantum Design PPMS equipped with an $^3$He refrigerator, down to a temperature of 0.35~K.

%%%%%%%%%%%%%%%%%%%%%%%%%%%%%%%%%%%%%%%%%%%%%%%%%%%%%%%%%%
\subsection{Crystalline electric fields excitations} %%%%%
%%%%%%%%%%%%%%%%%%%%%%%%%%%%%%%%%%%%%%%%%%%%%%%%%%%%%%%%%%

\begin{figure}[t]
    \centering
    \includegraphics{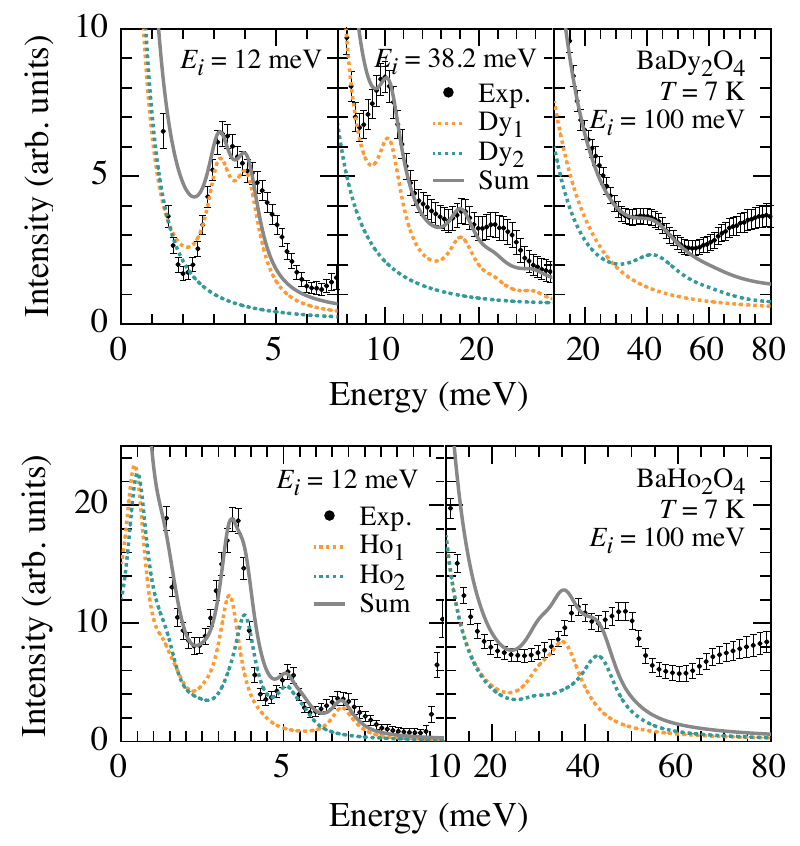}
    \caption[Inelastic neutron scattering spectra for \BDO{} and \BHO{} at $T$~=~7~K for various incident energies $E_i$]{Inelastic neutron scattering spectra for \BDO{} (top) and \BHO{} (bottom) at $T$~=~7~K for various incident energies $E_i$. Here, the spectra have been summed over all $|\mathbf{Q}|$ values. The solid line is the sum of the contributions from the two rare earth sites. The contribution from site 1 is shown as the orange dashed line and the contribution from site 2 is shown as the blue dashed line. The lines were obtained by fitting the experimental spectra with calculations using the program \textsc{multiX}.}
    \label{fig:Inelastic}
\end{figure}

The program \textsc{multiX} was used to calculate the inelastic neutron spectra for comparison with the experimental spectra. It computes the energy levels of an individual magnetic ion in the CEF defined by the charges at the positions of the neighboring ions. This permits the characterization of the single-ion anisotropy. Here, we are taking advantage of the fact that \textsc{multiX} allows us to calculate CEF levels in systems with low symmetry, such as the Ba\textit{Ln}$_2$O$_4$ structure, where the only point symmetry at the rare earth site is a mirror plane.

\begin{figure}[b]
    \centering
    \includegraphics{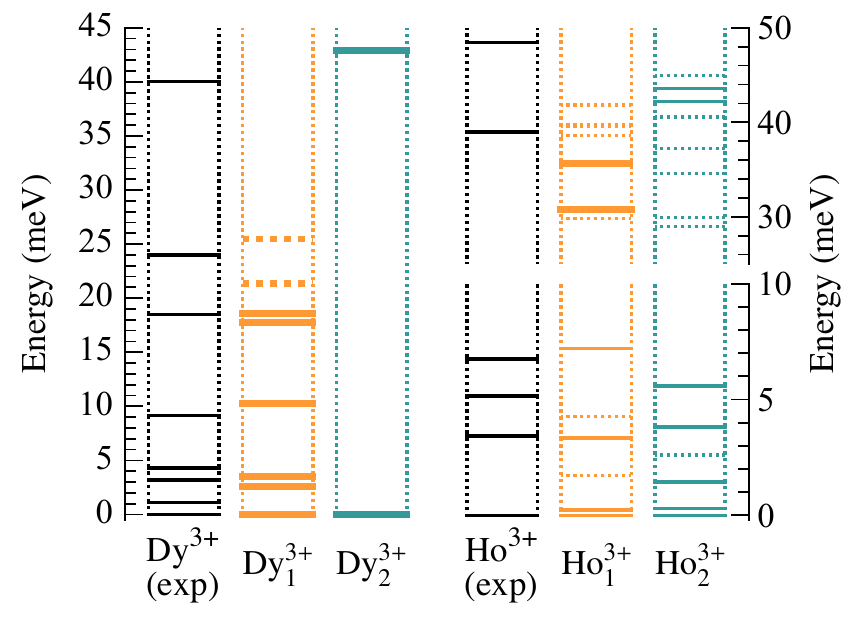} 
    \caption[The CEF level scheme of \BDO{} and \BHO{} as established by fitting theoretical calculations with \textsc{multiX} to inelastic neutron scattering data taken at $T$~=~7~K]{The CEF level scheme of \BDO{} (left) and \BHO{} (right) as established by fitting the levels from \textsc{multiX} to the inelastic neutron scattering spectra measured at $T$~=~7~K. The thick lines in the theoretical levels represent doublet or clear pseudo-doublet states, regular lines are singlet states, and dashed lines indicate levels of significantly lower intensity, which makes them hard to see in \hyperref[fig:Inelastic]{Fig.~\ref{fig:Inelastic}}. For a comparison with the calculations, Gaussian peaks were fitted to the visible peaks of the inelastic neutron scattering spectra, where the center of each Gaussian is shown as a line.}
    \label{fig:CEF_Levels}
\end{figure}

%The model is based on semi-empirical parameters scaling the intensities of the different components, due to the effect of the bonding environment of the ionized atom on the radial part of the wavefunction. We_ found that our_ systems are fairly insensitive in the Hamiltonian scaling of the spin-orbit interaction, which ats only locally on each atoms, asacting only on the wavefunction of a single site, and of the electrostatic interaction, taking into account wavefunctions overlap. 
Beside the CEF, the energy levels are calculated by \textsc{multiX} by taking into account the Coulomb and spin-orbit interaction. The strength of these three effects can be scaled in the model through semi-empirical parameters. The computed spectra are fairly insensitive to the scaling of the strength of the Coulomb and spin-orbit interaction of the Hamiltonian. In consequence, only the CEF scaling factor $S_{\mathrm{xtal}}$ was refined, resulting in a fair agreement with the experimental spectra, as shown in \hyperref[fig:Inelastic]{Fig.~\ref{fig:Inelastic}}. For these spectra the data for all wave vectors $|\mathbf{Q}|$ was summed. The calculations were carried out independently for the two crystallographically inequivalent rare earth sites, with a different CEF scaling factor for each site. In order to compare the calculations with the inelastic neutron spectra, the contribution from both sites was added. For \BDO{} the best fit was obtained with $S_{\mathrm{xtal}}^{\mathrm{Dy_1}}$~=~0.19 and $S_{\mathrm{xtal}}^{\mathrm{Dy_2}}$~=~0.64, and for \BHO{} with $S_{\mathrm{xtal}}^{\mathrm{Ho_1}}$~=~0.45 and $S_{\mathrm{xtal}}^{\mathrm{Ho_2}}$~=~0.58. These low values of the $S_{\mathrm{xtal}}$ parameters are similar to what has been observed in \SDHO{} \cite{Fennell:2014fy}, originating from the contraction of the {\textit f}-orbital, or an over-estimation of the ionic charges. An additional parameter was added to model the experimental broadening of the spectra. In the case of \BDO{}, the level of the first excited state is found at $E=$~2.6~meV and 42.9~meV for the Dy$^{3+}_1$ and Dy$^{3+}_2$ site, respectively, while for \BHO{} these levels are much closer to the ground states, with $E=$~0.2~meV and 0.3~meV for the Ho$^{3+}_1$ and Ho$^{3+}_2$, respectively. A comparison between the energy levels obtained by fitting the inelastic neutron scattering spectra and the position of the peaks in the spectra is shown in \hyperref[fig:CEF_Levels]{Fig.~\ref{fig:CEF_Levels}}. Here, the positions of the experimental levels were determined by fitting Gaussian functions to the peaks in the spectra.

\begin{table}
    \caption[Direction and size in $\mu_{\text{B}}$ of the magnetic moments as determined from crystal field calculations (CEF), using \textsc{multiX}, and from a refinement of the magnetic Bragg peaks of elastic neutron scattering (exp), for the two inequivalent rare earth sites in the \textit{AkLn}$_2$O$_4$ structure.]{Direction and size in $\mu_{\text{B}}$ of the magnetic moments as determined from crystal field calculations (CEF), using \textsc{multiX}, and from a refinement of the magnetic Bragg peaks of elastic neutron scattering (exp), for the two inequivalent rare earth sites in the \textit{AkLn}$_2$O$_4$ structure.}
    \label{table:Moments}
    \begin{ruledtabular} 
    \begin{tabular}{lcccclcccc}
    ~ & $\mu_{a}$ & $\mu_{b}$ & $\mu_{c}$ & $|\boldsymbol{\mu}|$ & ~ & $\mu_{a}$ & $\mu_{b}$ & $\mu_{c}$ & $|\boldsymbol{\mu}|$ \Bstrut\\  \hline \Tstrut
    Dy$_{\text{1}_{\text{CEF}}}$ & 9.4 & 2.4 & 0.1 & 9.7 & Dy$_{\text{1}_{\text{exp}}}$ & 2.9 & 4.2 & 0.0 & 5.1 \\
    Dy$_{\text{2}_{\text{CEF}}}$ & 1.9 & 9.6 & 0.0 & 9.8 & Dy$_{\text{2}_{\text{exp}}}$ & 0.0 & 2.3 & 0.0 & 2.3 \\ 
    Ho$_{\text{1}_{\text{CEF}}}$ & 1.2 & 9.6 & 0.0 & 9.7 & Ho$_{\text{1}_{\text{exp}}}$ & \multicolumn{4}{c}{Not ordered}   \\
    Ho$_{\text{2}_{\text{CEF}}}$ & 0.0 & 0.0 & 7.9 & 7.9 & Ho$_{\text{2}_{\text{exp}}}$ & 0.0 & 0.0 & 6.2 & 6.2 \\
    \end{tabular}
    \end{ruledtabular} 
\end{table}

% \begin{table}
%     \caption[]{}
%     \label{table:Moments}
%     \begin{ruledtabular} 
%     \begin{tabular}{lcclcc}
%     ~ & $\boldsymbol{\mu}$ & $|\boldsymbol{\mu}|$ & ~ & $\boldsymbol{\mu}$ & $|\boldsymbol{\mu}|$ \Bstrut\\  \hline \Tstrut
%     Dy$_{\text{1}_{\text{CEF}}}$ & $(9.4,2.4,0.1)$ & 9.7 & Dy$_{\text{1}_{\text{exp}}}$ & $(2.9,4.2,0.0)$ & 5.1 \\
%     Dy$_{\text{2}_{\text{CEF}}}$ & $(1.9,9.6,0.0)$ & 9.8 & Dy$_{\text{2}_{\text{exp}}}$ & $(0.0,2.3,0.0)$ & 2.3 \\ 
%     Ho$_{\text{1}_{\text{CEF}}}$ & $(1.2,9.6,0.0)$ & 9.7 & Ho$_{\text{1}_{\text{exp}}}$ & \multicolumn{2}{c}{Not ordered}   \\
%     Ho$_{\text{2}_{\text{CEF}}}$ & $(0.0,0.0,7.9)$ & 7.9 & Ho$_{\text{2}_{\text{exp}}}$ & $(0.0,0.0,6.2)$ & 6.2 \\
%     \end{tabular}
%     \end{ruledtabular} 
% \end{table}

The \textsc{multiX} calculations also predict the size and direction of the magnetic moments. Each Dy$^{3+}$ ion has a doublet ground state protected by time reversal symmetry due to the fact that Dy$^{3+}$ is a Kramers ion. For each Ho$^{3+}$, which is a non-Kramers ion, the ground state is treated as a pseudo-doublet because the lowest energy states are two singlets separated by an energy difference smaller than the computational accuracy. The moments obtained by this procedure are listed in \hyperref[table:Moments]{Table~\ref{table:Moments}}. For \BDO{}, the moments for both sites lie in the $ab$-plane, with one predominantly along the $a$-axis and the other along the $b$-axis. For \BHO{}, the first site has a moment in the $ab$-plane, while for the second site, the moment is along the $c$-axis. There is a clear distinction between moments laying exclusively in the $ab$-plane or exclusively along the $c$-axis --- perpendicular or along the chains. This points to a clear anisotropy of the magnetic interactions, as previously observed for the Sr variant of these compounds. We observe the following total moments per each site: $|\boldsymbol{\mu}_{\mathrm{Dy_{1}}}|~=~9.7\mu_{\text{B}}$, $|\boldsymbol{\mu}_{\mathrm{Dy_{2}}}|~=~9.8\mu_{\text{B}}$, $|\boldsymbol{\mu}_{\mathrm{Ho_{1}}}|~=~9.7\mu_{\text{B}}$, and $|\boldsymbol{\mu}_{\mathrm{Ho_{2}}}|~=~7.9\mu_{\text{B}}$. Of these, only the second Ho site has a moment which is significantly lower than the effective moment expected from Hund's rules --- 10.4$\mu_{\text{B}}$ for both Dy$^{3+}$ and Ho$^{3+}$ ions \cite{Marder:2010ta}. Fitting a Curie-Weiss model to the magnetic susceptibilities at high temperatures results in an effective moment $\mu_\mathrm{eff}$ per magnetic site of 10.60(1)$\mu_{\text{B}}$ and 10.77(1)$\mu_{\text{B}}$ for \BDO{} and \BHO{}, respectively, with Curie-Weiss temperatures $\theta_{\mathrm{CW}}$ of $-$18.5(3)~K and $-$10.9(3)~K.

\begin{figure}
    \centering
    \includegraphics{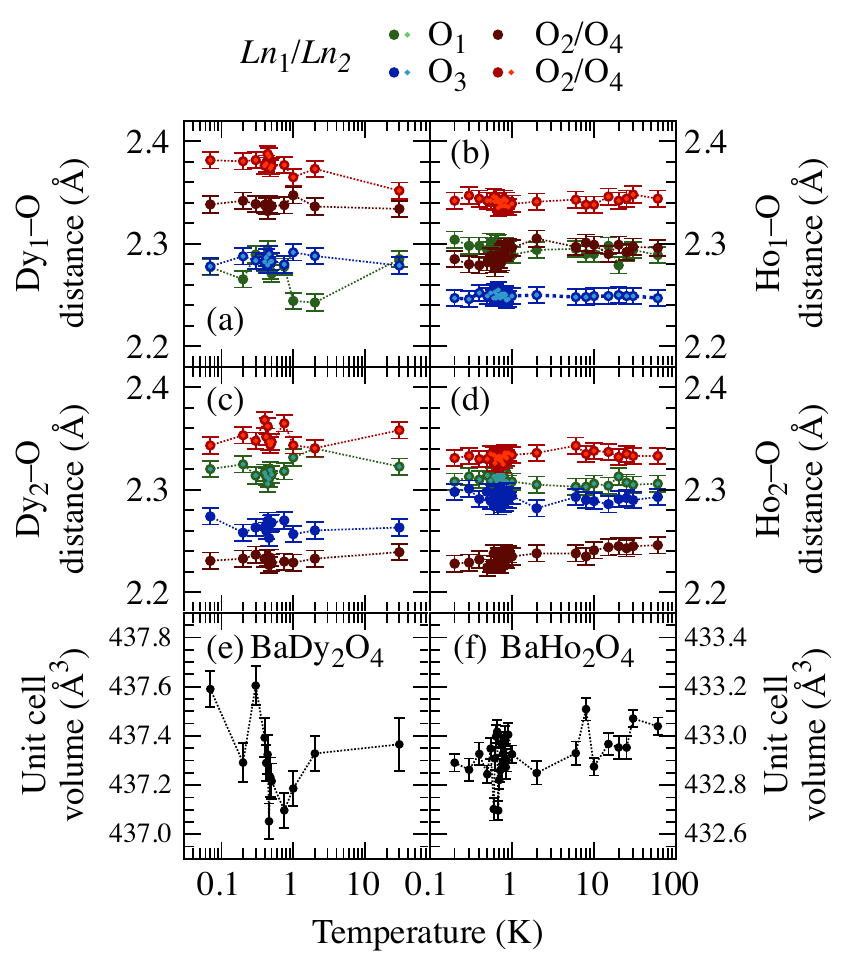} 
    \caption[Temperature dependence of the crystal structure from a refinement of the elastic neutron scattering data]{Temperature dependence of the crystal structure from a refinement of the elastic neutron scattering data. Figures (a)--(d) show the distance between the rare earth atoms and the six surrounding O atoms, which is constant within the resolution of our experiment. The labels on the O atoms represent their four different positions in this crystal structure. Figures (e)--(f) present the very small variation of the unit cell volume.}
    \label{fig:Cell_Param}
\end{figure}

The calculations of the CEF levels with \textsc{multiX} are fairly sensitive to the details of the crystallographic structure. The low symmetry of the structure is the source of this complex level scheme, and slight distortions of the structure could have a great impact on the results, as this calculation is highly dependent on the atomic positions. It is then important to determine whether any significant distortion occurs upon cooling. For this, we refined the lattice constants and the atomic positions within the unit cell from the neutron diffraction spectra at high angles 2$\theta$, as the magnetic scattering intensity is stronger at low angles. For \BHO{}, a simultaneous refinement of the data acquired at wavelengths $\lambda$ of 1.886~\AA{} and 1.155~\AA{} was performed. Only a small variations of the O distances is observed in the entire range of temperatures from 60~K down to 0.07~K, as shown in \hyperref[fig:Cell_Param]{Fig.~\ref{fig:Cell_Param}}. The chemical unit cell volume varies by about 0.5~\AA$^3$ for \BDO{}, and there is no volume change for \BHO{} within the experimental accuracy. The slight change in the lattice parameters of \BDO{} is accompanied by a small increase of the Bragg peaks' width. The intra-chain distance between the rare earth atoms is also temperature independent and not affected by the magnetic order. The dimension of the O octahedra surrounding the magnetic ions have the biggest influence on the CEF splitting, and a thermal distortion would affect the single-ion anisotropy. However, no such distortion is observed for both \BDO{} and \BHO{}. Such a distortion was previously noted in the structurally equivalent compounds of SrTb$_2$O$_4$ and SrTm$_2$O$_4$ \cite{Li:2014gs,Li:2015dh}, but the change in these compounds is rather subtle and below our experimental resolution. For \BDO{}, the lattice parameters are $a$~=~10.4068(3) \AA{}, $b$~=~12.1131(3) \AA{} and $c$~=~3.46985(10) \AA{} at $T$~=~0.07~K, and for \BHO{}, $a$~=~ 10.3864(3) \AA, $b$~=~12.0852(2) \AA{} and $c$~=~3.44754(8) \AA{} at $T$~=~0.20~K. These values are similar to those reported by Doi {\textit{et al.}} \cite{Doi:2006ue}.

\begin{figure} 
    \centering
    \includegraphics{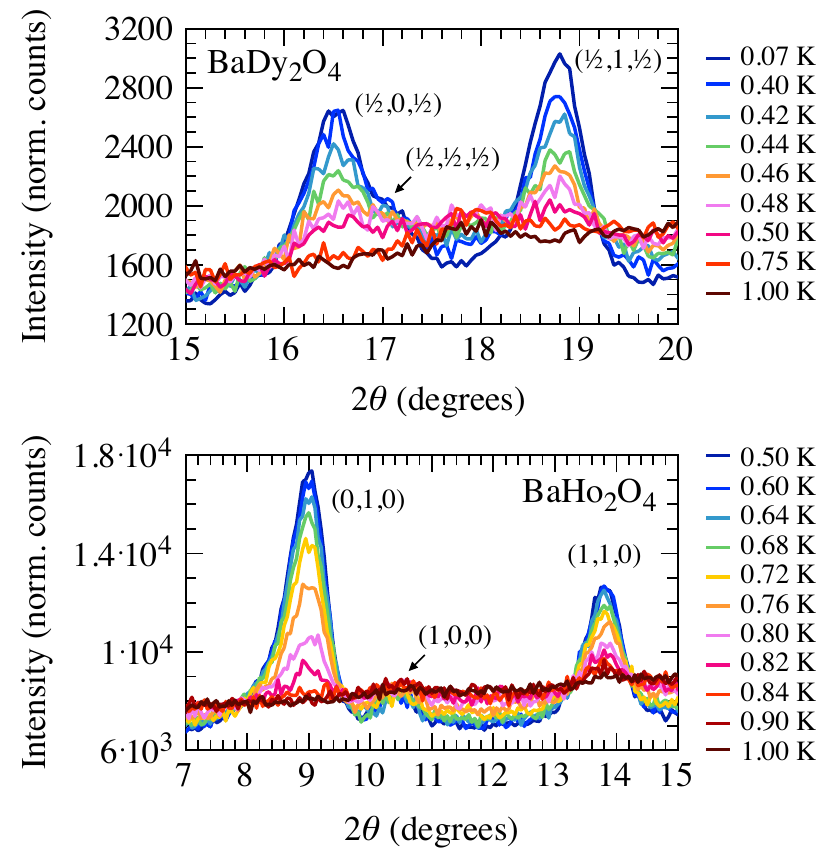} 
    \caption[Thermal evolution of the magnetic Bragg peaks at low diffraction angles]{Thermal evolution of the magnetic Bragg peaks at low diffraction angles. The top panel shows the magnetic peaks of \BDO{} at {\bf{Q}} of ($\frac{1}{2}$,0,$\frac{1}{2}$), ($\frac{1}{2}$,$\frac{1}{2}$,$\frac{1}{2}$) and ($\frac{1}{2}$,1,$\frac{1}{2}$), indicating the presence of two propagation vectors, $\mathbf{k}_1$~=~($\frac{1}{2}$,0,$\frac{1}{2}$) and $\mathbf{k}_2$~=~($\frac{1}{2}$,$\frac{1}{2}$,$\frac{1}{2}$). The data for \BHO{} is shown on the lower panel, were magnetic peaks appear at $\mathbf{k}$~=~(0,0,0) values of {\bf{Q}}.}
    \label{fig:Magnetic_Peaks}
\end{figure}

\begin{figure} 
    \centering
    \includegraphics{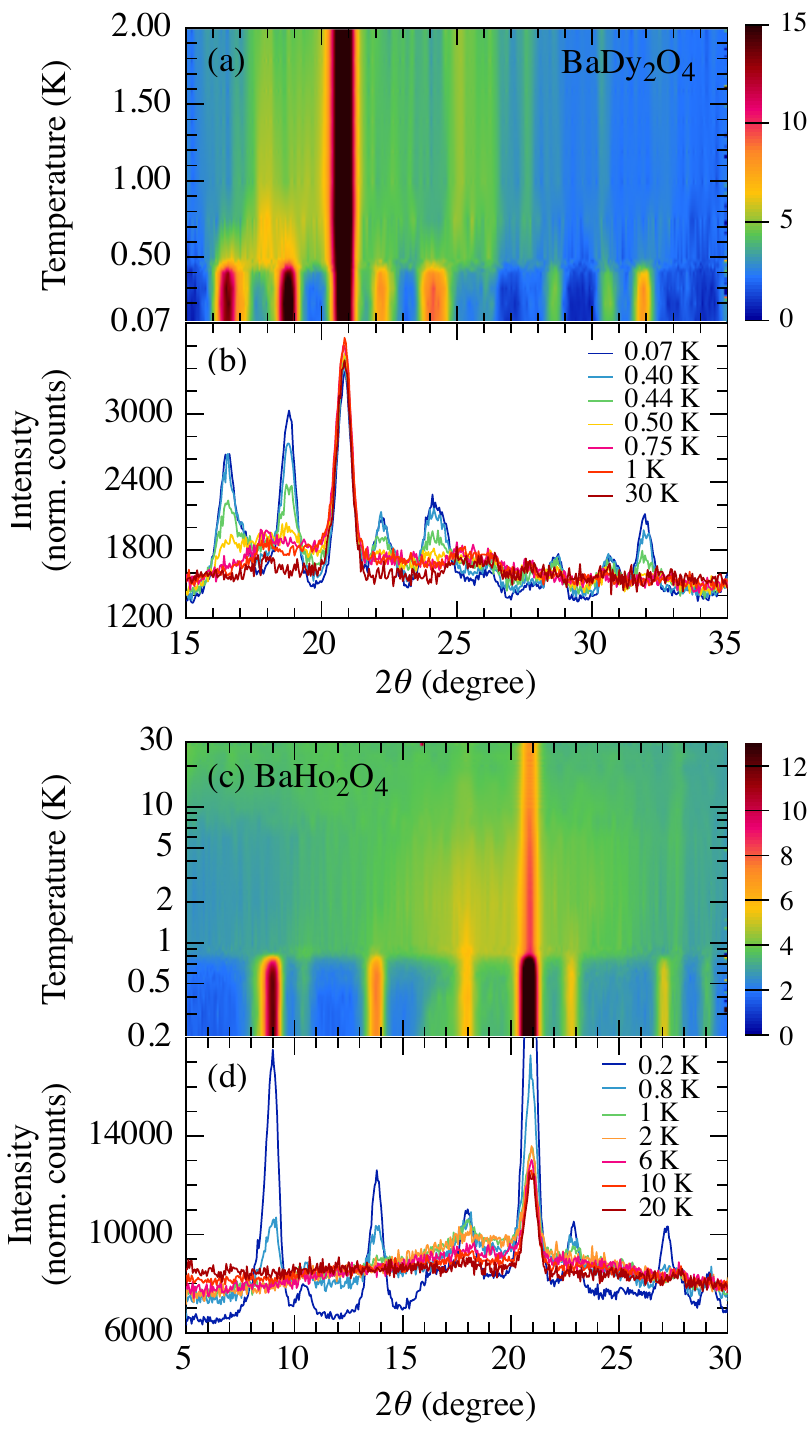} 
    \caption[Principal magnetic Bragg peaks with emphasis on the thermal evolution of the diffuse scattering seen on the two-dimensional color plots]{Region showing the principal magnetic Bragg peaks to illustrate the thermal evolution of the diffuse scattering. The diffusive scattering is clearly visible in the two-dimensional color plots. Figures (a)--(b) show a sawtooth feature for \BDO{}, which is centered at $2\theta\sim18\degree$ just above the transition. Figures (c)--(d) present a similar feature for \BHO{}, also centered at $2\theta\sim18\degree$, which persists to temperatures below the ordering temperature of 0.84~K.}
    \label{fig:2D_Spectra}
\end{figure}

The CEF level scheme of \BDO{} was robust to repositioning the atoms within the error bars of our structural refinement and only a scaling of the level separation was observed. However, in \BHO{}, the level scheme was more strongly affected, and significantly changes when the O positions are set to the limit of the uncertainties of our refinement. Despite the absence of clear structural changes, the \textsc{multiX} calculations would nonetheless benefit from higher resolution data from a X-ray synchrotron source to observe potential distortion of the O octahedra. Additional measurements are planned, such as electron paramagnetic resonance (EPR) and photoluminescence spectroscopy, as were used in the case of SrEr$_2$O$_4$ \cite{Malkin:2015eh} to determine the CEF levels.

The motivation for carrying out these CEF calculations was to establish a basis for the characterization of the magnetic structures in the following sections. The results from the CEF model for the magnetic anisotropy and direction of the easy-axis are important for developing a magnetic model.

% For now, the interest in the CEF calculations was mainly to lay down a backbone for the following characterization of the magnetic structure through the analysis of the elastic neutron powder diffraction. Hints concerning the magnetic anisotropy and direction of the easy-axis will prove to be valuable informations on discerning the interplay between interactions.

% \textcolor{TealBlue}{As opposed to what as been claimed in the Sr(Dy,Ho)2O4's paper, the Spin Orbit scaler do affect the fit a lot. It is possible to get extremely good fit while changing this parameter. Is there a physical argument to prioritize CEF scaler over the spin-orbit scaler ? Current calculations are produced with SO = Coulomb = 0.8, as in the default file. Since there should not be structural change with temperature, only one scheme is produce. The code also have a distinction between CEF scaler of all electrons, core electrons and valence electrons. Should we_ care about only the valence ? Probably yes.}

%%%%%%%%%%%%%%%%%%%%%%%%%%%%%%%%%%%%%
\subsection{Magnetic structure} %%%%%
%%%%%%%%%%%%%%%%%%%%%%%%%%%%%%%%%%%%%

\begin{figure} 
    \centering
    \includegraphics{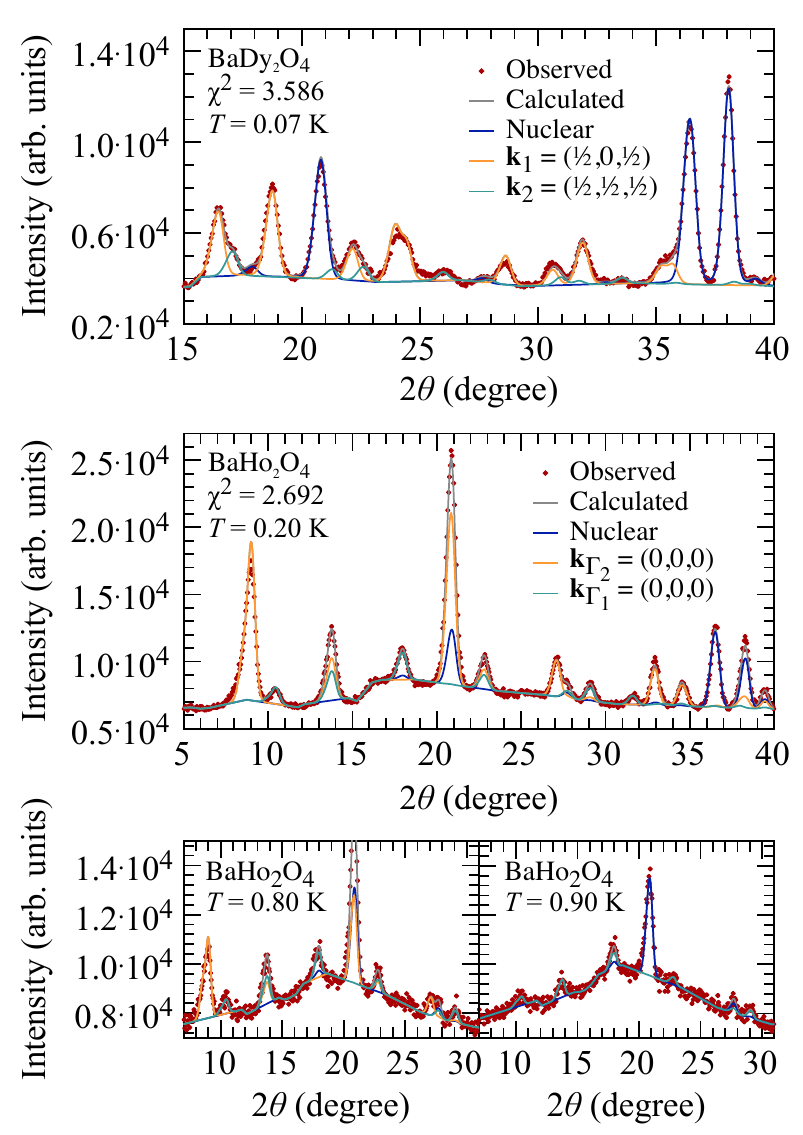} 
    \caption[FullProf refinement of the neutron diffraction patterns of \BDO{} and \BHO{} taken at $T=$~0.07~K and 0.20~K, respectively]{FullProf refinement of the neutron diffraction patterns of \BDO{} (top) and \BHO{} (bottom) taken at $T=$~0.07~K and 0.20~K, respectively. The nuclear and the different magnetic contributions are plotted separately. In \BDO, the magnetic contributions come from two wave vectors $\mathbf{k}_1$~=~($\frac{1}{2}$,0,$\frac{1}{2}$) and $\mathbf{k}_2$~=~($\frac{1}{2}$,$\frac{1}{2}$,$\frac{1}{2}$). In \BHO{}, the magnetic contributions have a single wave vector $\mathbf{k}_0$~=~(0,0,0), but two different irreducible representations, $\Gamma_1$ and $\Gamma_2$. This results in two magnetic transitions with different critical temperatures. The lowest panels shows the refinement of the \BHO{} data close to the lower transition, where a new set of Bragg peaks emerges as the temperature is lowered from 0.9~K to 0.8~K.}
    \label{fig:Fit_FullProf}
\end{figure}

%%%%%%%%%%%%%%%%%%%%%%%%%%%%%%%%%%%%%%%%%%%%%%%%%%%%%%%%%
\subsubsection{\texorpdfstring{\BDO}{}} %%%%%
%%%%%%%%%%%%%%%%%%%%%%%%%%%%%%%%%%%%%%%%%%%%%%%%%%%%%%%%%

Information related to the ordering temperature as well as the nature of the magnetic order can be deduced from the temperature dependence of the powder diffraction spectra. Spectra at low diffraction angles $2\theta$ from base temperature to $T$~=~1~K are shown in \hyperref[fig:Magnetic_Peaks]{Fig.~\ref{fig:Magnetic_Peaks}}. In the case of \BDO{}, upon cooling below $T$~=~0.48~K, additional Bragg peaks appear at $2\theta$'s of $16.5\degree$, $17.1\degree$, and $18.8\degree$. These peaks can be indexed with a momentum transfer $\mathbf{Q}$ of ($\frac{1}{2}$,0,$\frac{1}{2}$), ($\frac{1}{2}$,$\frac{1}{2}$,$\frac{1}{2}$), and ($\frac{1}{2}$,1,$\frac{1}{2}$), which indicate a magnetic order with two propagation vectors $\mathbf{k}_1$~=~($\frac{1}{2}$,0,$\frac{1}{2}$) and $\mathbf{k}_2$~=~($\frac{1}{2}$,$\frac{1}{2}$,$\frac{1}{2}$).

\begin{figure*}[t] 
    \centering
    \includegraphics[width=0.9\textwidth]{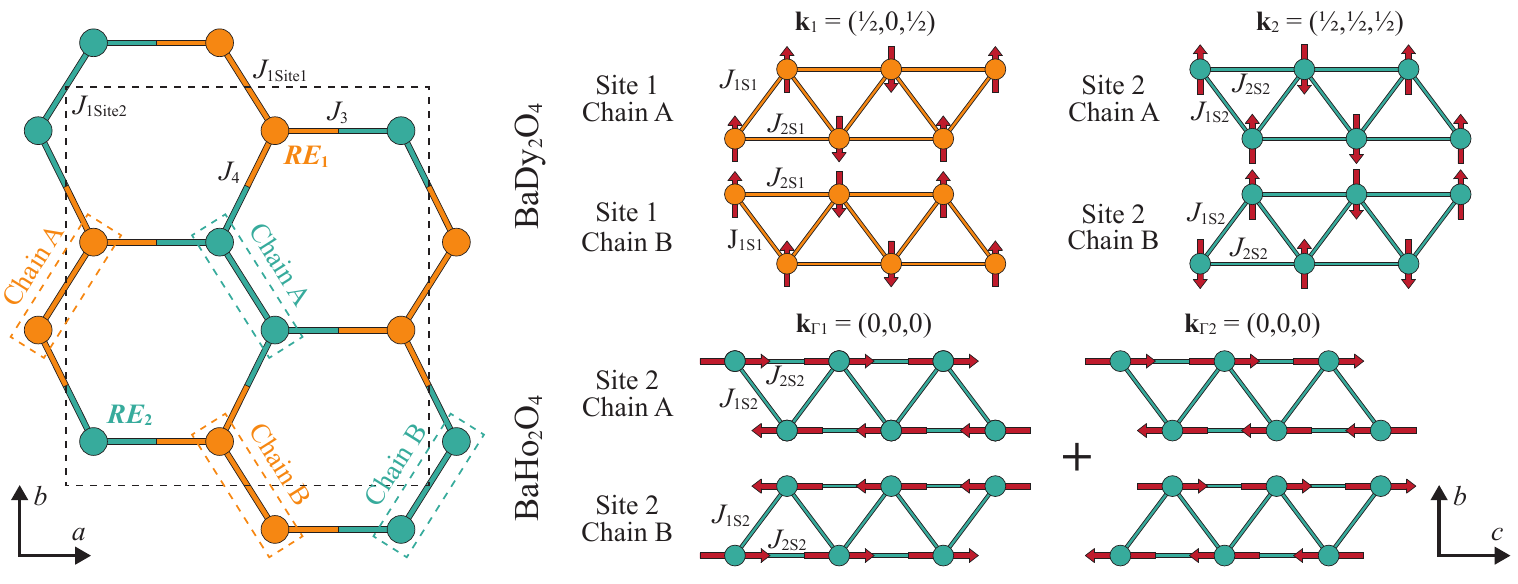} 
    \caption[Arrangement of the rare earth atoms and magnetic structure of \BDO{} and \BHO{}]{Arrangement of the rare earth atoms and magnetic structure of \BDO{} and \BHO{}. The left side illustrates a view of the $ab$-plane, exhibiting the honeycomb pattern of the rare earth atoms, where the dotted box is a chemical unit cell. For a clearer view, Ba and O atoms were omitted, and the crystallographically non-equivalent rare earth sites, which each consist of two chains stretching along the $c$-axis, are shown in a different color. $J_1$ and $J_2$ are the interaction constants for the nearest and next-nearest neighbors along the chains, while $J_3$ and $J_4$ are the interaction constants between nearest neighbors outside the chains, located in direction of $a$ and $b$, respectively. On the right side of the figure, the chains along the $c$-axis are isolated and projected onto the $bc$-plane. In \BDO{}, both ordering vectors represent a double N\'eel arrangement. For $\mathbf{k}_1$, the two chains A and B stay equivalent. However, for $\mathbf{k}_2$, the moments on chain B are shifted by one position with respect to their positions on chain A. Only a single transition temperature is observed, for both wave vectors. In \BHO{}, site 2 exhibits a simple N\'eel order, which is split into two different irreps. Each irrep has its own ordering temperature while the moments on site 1 do not order.}
    \label{fig:Structure}
    %For \BDO{}, the two propagation vectors were separated on different sites for the purpose of clarity, despite the fact that they may coexist on the same site.
\end{figure*}

%For \BHO{}, new peaks are observed at integer values of {\bf Q} for temperatures below 0.84~K, as well as additional intensity at the nuclear peaks positions at higher diffraction angles, indicating a magnetic order with a propagation vector {\bf k}~=~(0,0,0).

On a wider range of angles, as shown in \hyperref[fig:2D_Spectra]{Fig.~\ref{fig:2D_Spectra}(a)--(b)}, one can notice the evolution of an additional magnetic scattering signal when the magnetic transition temperature is approached. This additional signal has the form of a sawtooth, which is reminiscent of the powder average from one-, or two-dimensional magnetic correlations \cite{Warren:1941cj,Jones:1949cx}. This diffuse scattering is present in both compounds, and contrary to what one would expect for a fully ordered system, it does not completely disappear with the onset of the magnetic Bragg peaks, but remains visible down to the lowest measured temperatures. The main feature of the diffuse scattering is located between the Bragg peaks at $\mathbf{Q}$~=~($\frac{1}{2}$,$\frac{1}{2}$,$\frac{1}{2}$) and ($\frac{1}{2}$,1,$\frac{1}{2}$), at $2\theta\sim18\degree$. It is outlined by the strong scattering intensity in the color plot and clearly visible up to at least $T=2$~K. Below the transition, this feature is harder to see due to high density of magnetic Bragg peaks, but it persists down to the lowest measured temperatures. The color plot ends at $T=2$~K, as the next pattern at a higher temperature was only taken at 30~K.

%The result of the refinement of the diffraction patterns is shown in Fig.~\ref{fig:Fit_FullProf}. It provides a comprehensive picture of the magnetism present in these systems --- both possess two distinct magnetic orders of different nature.
The result of the refinement of the \BDO{} diffraction pattern is shown in the top panel of \hyperref[fig:Fit_FullProf]{Fig.~\ref{fig:Fit_FullProf}}. The natural solution for this magnetic order is that the two different ordering vectors are each associated with a crystallographically different rare earth site. The sharpness of the magnetic Bragg peaks indicates that the order is long-range, as their widths are comparable to the nuclear Bragg peaks. A visualization of the structures is given in \hyperref[fig:Structure]{Fig.~\ref{fig:Structure}}.

Each of the two magnetic sites forms a chain which has a double N\'eel structure with a $\uparrow\uparrow\downarrow\downarrow$ motif. As noted in \hyperref[fig:Magnetic_Peaks]{Fig.~\ref{fig:Magnetic_Peaks}}, site 1 has a propagation vector $\mathbf{k}_1$~=~($\frac{1}{2}$,0,$\frac{1}{2}$) and site 2 has $\mathbf{k}_2$~=~($\frac{1}{2}$,$\frac{1}{2}$,$\frac{1}{2}$), quadrupling and octupling the volume of the chemical unit cell. Both ordering vectors correspond to the same Shubnikov group with different irreducible representations (irreps) of the space group {\textit{Pnam}}, as listed in \hyperref[table:irreps]{Table~\ref{table:irreps}}. Chain A has a $\uparrow\uparrow\downarrow\downarrow$ motif for both sites. In contrast, on chain B, one site has a $\uparrow\uparrow\downarrow\downarrow$ motif, the other a $\downarrow\uparrow\uparrow\downarrow$ motif.

The structure was refined for all measured temperatures, which made it possible to extract the temperature dependence of the size and direction of the magnetic moments. For this analysis, the scaling parameters in the refinement were fixed to the one from the nuclear Bragg peaks intensity, in order to obtain the correct moment size. The results for both propagation vectors and irreps are shown in \hyperref[fig:Moments_size_vs_temperature]{Fig.~\ref{fig:Moments_size_vs_temperature}}.

%We now describe the temperature dependence, as well as the direction of the magnetic moments. The intensity of the different magnetic peaks is proportional to the square of the magnetic moment associated with the magnetic order. Here, the scaling parameters were fixed to the ones from nuclear Bragg peaks intensity. We were able to follow the temperature evolution of the individuals orders associated with their respective propagation vectors and irreps, as shown in \hyperref[fig:Moments_size_vs_temperature]{Fig.~\ref{fig:Moments_size_vs_temperature}}. At low temperatures, the intensity of the observed magnetic orders saturates for all of them.

From the refinement, we know that both moments are laying in the $ab$-plane: $\boldsymbol{\mu}_{\mathbf{k}_1}~=~(2.9,4.2,0.0)\mu_{\text{B}}$ and $\boldsymbol{\mu}_{\mathbf{k}_2}~=~(0.0,2.3,0.0)\mu_{\text{B}}$. Also, both moments order simultaneously at $T_\mathrm{N} = 0.48$~K, within the temperature resolution of our experiment, and they saturate already by 0.30~K.

From the analysis of the powder spectra, it is not possible to assign the different wave vectors to a particular rare earth site --- the $\chi^2$ from the refinements are not significantly different when we exchange the sites, or if both wave vectors are confined to a single site. Nonetheless, identification of the sites is possible by comparing the direction of the magnetic moments obtained from the CEF calculations for the different crystallographic sites, listed in \hyperref[table:Moments]{Table~\ref{table:Moments}}, with the direction of the moments from the refinement of the powder spectra. For \BDO{}, the \textsc{multiX} calculations and the refinement from the experimental diffraction pattern both show that both moments are restricted to the $ab$-plane. We can now use the relative strength of the components along $a$ and $b$ to distinguish between the two sites --- the moment $\boldsymbol{\mu}_{\mathrm{Dy_1}}$ is stronger along $a$ and the moment $\boldsymbol{\mu}_{\mathrm{Dy_2}}$ stronger along $b$. The resulting magnetic structure is shown in \hyperref[fig:Structure]{Fig.~\ref{fig:Structure}}, where we show the different sites and chains separately for clarity. The discrepancies between the size of the magnetic moments from the refinements and those predicted by the CEF calculations can be explained by the presence of diffuse scattering, analyzed in \hyperref[sec:diffuse]{Sec.~\ref{sec:diffuse}}.
%while keeping overall scaling factor constant

%%%%%%%%%%%%%%%%%%%%%%%%%%%%%%%%%%%%%%%%%%%
\subsubsection{\texorpdfstring{\BHO}{}} %%%
%%%%%%%%%%%%%%%%%%%%%%%%%%%%%%%%%%%%%%%%%%%

\begin{table*}
    \caption[Basis vectors of the irreducible representations of the space group {\textit{Pnam}} (No. 62), where the magnetic ions occupy position 4$c$ which has a $m$ symmetry, for each propagation vector {\bf k} of the magnetic structures of \BDO{} and \BHO{}.]{Basis vectors of the irreducible representations of the space group {\textit{Pnam}} (No. 62), where the magnetic ions occupy position 4$c$ which has a $m$ symmetry, for each propagation vector {\bf k} of the magnetic structures of \BDO{} and \BHO{}.}
    \label{table:irreps}
    \begin{ruledtabular}
    \begin{tabular}{lll}
    \BDO{} & $\mathbf{k}_1$ = ($\frac{1}{2}$,0,$\frac{1}{2}$) & $\mathbf{k}_2$ = ($\frac{1}{2}$,$\frac{1}{2}$,$\frac{1}{2}$)            \Bstrut \\ \hline
    x,y,z                                             & $(C_1-iC_4, C_2-iC_5, C_3+iC_6)$  & $(C_1-iC_4, C_2-iC_5, C_3+iC_6)$      \Tstrut \\ \hline
    -x,-y,z+$\frac{1}{2}$                             & $(-C_4+iC_1,-C_5+iC_2,-C_6-iC_3)$ & $(-C_4+iC_1,-C_5+iC_2,-C_6-iC_3)$  \\ \hline
    x+$\frac{1}{2}$,-y+$\frac{1}{2}$,-z+$\frac{1}{2}$ & $(-C_4-iC_1,C_5+iC_2,-C_6+iC_3)$  & $(C_4+iC_1,-C_5-iC_2,-C_6+iC_3)$  \\ \hline
    -x+$\frac{1}{2}$,y+$\frac{1}{2}$,-z               & $(C_1+iC_4, -C_2-iC_5, C_3-iC_6)$ & $(C_1+iC_4, -C_2-iC_5, -C_3+iC_6)$ \\ \hline\hline
    \BHO{}                                            & $\mathbf{k}_{\Gamma_1}$ =  (0,0,0)            & $\mathbf{k}_{\Gamma_2}$ = (0,0,0)                 \Bstrut \Tstrut \\ \hline
    x,y,z                                             & $(0,0,C_1)$                       & $(0,0,C_1)$                           \Tstrut \\ \hline
    -x,-y,z+$\frac{1}{2}$                             & $(0,0,-C_1)$                      & $(0,0,-C_1)$                       \\ \hline
    x+$\frac{1}{2}$,-y+$\frac{1}{2}$,-z+$\frac{1}{2}$ & $(0,0,C_1)$                       & $(0,0,-C_1)$                       \\ \hline
    -x+$\frac{1}{2}$,y+$\frac{1}{2}$,-z               & $(0,0,-C_1)$                      & $(0,0,C_1)$                        \\ 
    \end{tabular}
    \end{ruledtabular}
\end{table*}

For \BHO{}, new peaks are observed for temperatures below $T=0.84$~K at values of {\bf Q} corresponding to a propagation vector $\mathbf{k}_0$~=~(0,0,0). They are clearly visible at low angles in \hyperref[fig:Magnetic_Peaks]{Fig.~\ref{fig:Magnetic_Peaks}}. Additional intensity also appears at the position of nuclear Bragg peaks at higher diffraction angles.

The diffusive scattering in \hyperref[fig:2D_Spectra]{Fig.~\ref{fig:2D_Spectra}(c)--(d)} is similar to the case of \BDO{}. However, it is easier to see in \BHO{} with its $\mathbf{k}_0$ propagation vector, where once again the diffusive signal is stronger at $2\theta\sim18\degree$. This diffuse scattering signal is still visible on the otherwise stable background up to $T=30$~K, but also persist below the magnetic transition down to 0.20~K where the feature becomes much sharper.

The result of the refinement of the diffraction patterns is shown in the lower panels of \hyperref[fig:Fit_FullProf]{Fig.~\ref{fig:Fit_FullProf}}. Like its sister compound \BDO{}, \BHO{} possesses two distinct magnetic orders, characterized by sharp resolution limited magnetic Bragg peaks. In this case, they consist of simple N\'eel chains $\uparrow\downarrow\uparrow\downarrow$ with a propagation vector $\mathbf{k}_0$~=~(0,0,0), which does not increase the size of the unit cell. Two magnetic orders with different irreps, $\Gamma_1$ and $\Gamma_2$, are nonetheless needed in order to adequately account for magnetic scattering intensities. With such propagation vector, each possible irrep is forced by symmetry to either carry a moment in the $ab$-plane or along the $c$-axis, whereas the irreps of $\mathbf{k}_1$ and $\mathbf{k}_2$ of \BDO{} are not restraining the moment direction.

Moreover, a close inspection of the data together with the refinement, presented in the lower panels of \hyperref[fig:Fit_FullProf]{Fig.~\ref{fig:Fit_FullProf}}, show that the two orders --- or sets of Bragg peaks --- appear in two steps. The contribution from $\Gamma_1$ indicates order at 0.84~K, and the one from $\Gamma_2$ orders between 1~K and 2~K, a temperature range in which unfortunately no additional was gathered. Both orders share the same propagation vector and have ordered moments perpendicular to the $ab$-plane, pointing along $c$. Interestingly, this points to a coexistence of the ordered moments on site 2, leaving the moment on site 1 disordered. This is then consistent with the \textsc{multiX} calculations which show that the moment of the Ho on site 2, $\boldsymbol{\mu}_{\mathrm{Ho_2}}$, points along $c$ while the moment on site 1, $\boldsymbol{\mu}_{\mathrm{Ho_1}}$, is restricted to the $ab$-plane. There are several possibilities to interpret this structure from the powder diffraction refinement. One rather unusual interpretation is to treat the two orders as distinct phases. This would imply the superposition of the two orders on the same site, resulting in magnetic moments on chains A and B with starkly different sizes.

\begin{figure} 
    \centering
    \includegraphics{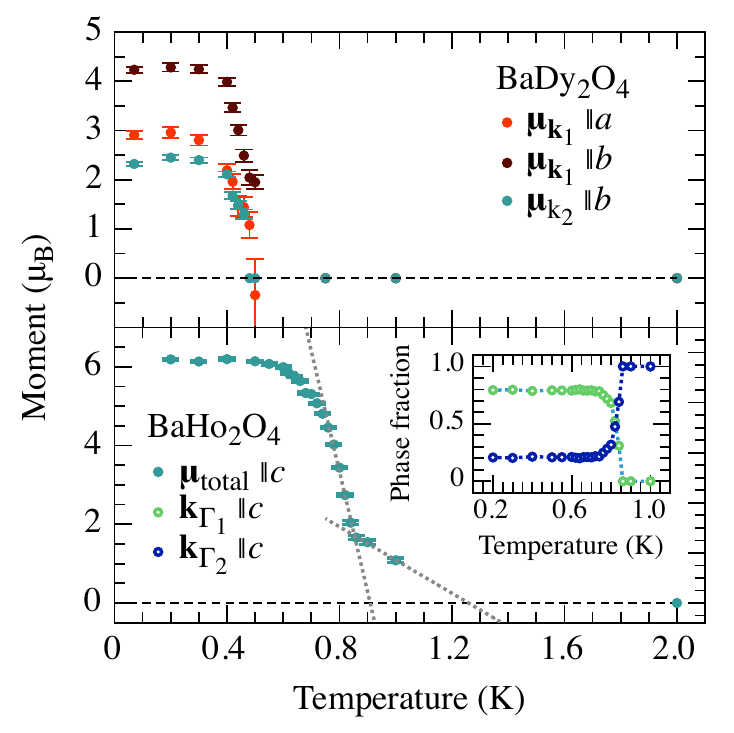} 
    \caption[Size and direction of the magnetic moments as determined from a refinement of the neutron powder diffraction data]{Size and direction of the magnetic moment as determined from the refinement of the neutron powder diffraction data. In \BDO{}, the moments for both wave vectors $\mathbf{k}_1$ and $\mathbf{k}_2$ order simultaneously at $T_\mathrm{N} = 0.48$~K. In \BHO{}, the two irreps, which have the same wave vector $\mathbf{k}_0 =(0,0,0)$, order at different temperatures. The dotted lines are linear extrapolations. The inset shows the phase fraction of the two irreps, where $\Gamma_2$ orders at approximately $T_{\mathrm{N}\Gamma_2}\sim1.3$~K, and $\Gamma_1$ at $T_{\mathrm{N}\Gamma_1} = 0.84$~K. At the lowest temperatures, the ratio between the two irreps stabilizes at 80\%--20\%.}
    \label{fig:Moments_size_vs_temperature}
\end{figure}

The more physical scenario is that the correlations in the chains nucleate either one of the two magnetic structures, leading to a phase separation within a single chain. In this case, the same moment is placed on both irreps, while the phase fraction is allowed to vary. The refinement of \BHO{} yields a total moment $\boldsymbol{\mu}_{\mathbf{k}_0}~=~(0.0,0.0,6.2)\mu_{\text{B}}$, where the moment increases with decreasing temperature in two steps, as seen in the lower panel of \hyperref[fig:Moments_size_vs_temperature]{Fig.~\ref{fig:Moments_size_vs_temperature}}. The inset shows the phase fraction of each irrep, normalized to 1. From $T_{\mathrm{N}\Gamma_2}\sim1.3$~K down to the second ordering temperature $T_{\mathrm{N}\Gamma_1}=0.84$~K, only $\Gamma_2$ contributes to the magnetic diffraction. At this temperature, a new set of Bragg peaks characterized by $\Gamma_1$ appears. For temperatures below the onset of the new Bragg peaks, near $T$~=~0.75~K, the phase fraction stabilizes at a ratio 80\% of $\Gamma_1$ and 20\% of $\Gamma_2$. This conclusion is also supported by our analysis of the diffuse scattering intensity, detailed in the next section.% which is present down to the lowest measured temperatures.

\begin{figure} 
    \centering
    \includegraphics{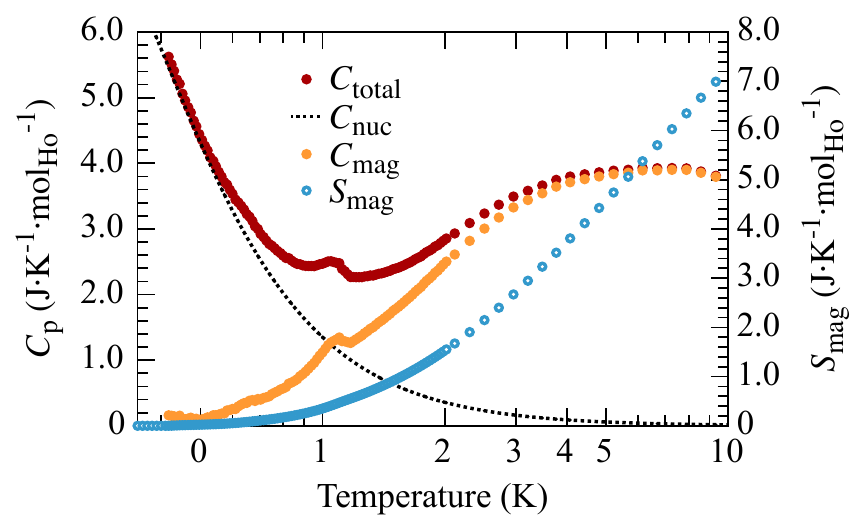} 
    \caption[Specific heat measurement \BHO{} on single crystals as a function of $T$]{Specific heat measurement of \BHO{} on single crystals as a function of $T$. The dark red points are the measured specific heat $C_\mathrm{total}$, the dashed line the calculated nuclear contribution $C_\mathrm{nuc}$, and the orange points the magnetic contribution $C_\mathrm{mag}$, resulting from the subtraction of $C_\mathrm{nuc}$ from $C_\mathrm{total}$. A small anomaly is located around $T=1.2$~K, coherent with the magnetic transition observed in neutron powder diffraction. The blue circles are the magnetic entropy $S_\mathrm{mag}$, calculated from $\int C_\mathrm{mag}(T)/T \mathrm{d}T$, and associated with the right-sided axis.}
    \label{fig:Specific_Heat}
\end{figure}

Previous specific heat measurements on powder of \BHO{} report only one antiferromagnetic phase transition at $T_\mathrm{N}=0.8$~K \cite{Doi:2006ue}. This is in contradiction with our own measurements which show a transition at $T_\mathrm{N}$~=~1.2~K, which is close to $T_{\mathrm{N}\Gamma_2}\sim1.3$~K, associated with the linear extrapolation of the moment size in \hyperref[fig:Moments_size_vs_temperature]{Fig.~\ref{fig:Moments_size_vs_temperature}}. The specific heat measurements shown in \hyperref[fig:Specific_Heat]{Fig.~\ref{fig:Specific_Heat}} were performed on small single crystals obtained through flux growth \cite{Besara:2014kx}. The large increase seen at the lowest temperatures is the signature of a nuclear Schottky contribution $C_\mathrm{nuc}$. Subtracting $C_\mathrm{nuc}$ from the measured data yields the magnetic contribution $C_\mathrm{mag}$, as in this temperature range, the lattice contribution is negligible. The entropy $S_\mathrm{mag}$ is calculated by integrating $C_\mathrm{mag}(T)/T$. While the temperature range is too narrow to show a saturation, it is interesting at this point to note that the magnetic transition has only a very small contribution to the entropy of the system. Just above $T_\mathrm{N}$~=~1.2~K, it is well below the expected value of $R\ln(2)$ for the ordering of a doublet, where $R$ is the universal gas constant. This indicates very little difference in entropy between the fluctuating chains and long-range magnetic order.

% This conclusion is also supported by our_ analysis of the diffuse scattering intensity, detailed in the next section. Its presence down to the lowest measured temperatures could also explain the discrepancies with the magnetic intensities predicted by the CEF calculations.

% \textcolor{TealBlue}{At this point in the article, we_ cannot assign site 1 and site 2. Scenario will come with the diffuse scattering analysis. For \BHO, it is clear from the diffuse scattering that these two orders are on the same site and the other site is in the $ab$-plane and not ordering. The intensity of the moment in the $ab$-plane would be about 8.5 $\mu_\mathrm{b}$ and mostly temperature independent, which is quite strong, and coherent with the CEF calculations.}

%%%%%%%%%%%%%%%%%%%%%%%%%%%%%%%%%%%%%%%%%%%%%%%%%%%%%%%%%%%%%%%%%
\subsection{1D magnetic correlations in diffuse scattering}\label{sec:diffuse}%%%%%
%%%%%%%%%%%%%%%%%%%%%%%%%%%%%%%%%%%%%%%%%%%%%%%%%%%%%%%%%%%%%%%%%

% The shape of the diffuse scattering is similar to what has been observed in \SDO{} and \SHO{}. Our_ sawtooth spectra are smooth, indication that the interactions are mainly 1D, as two- or three-dimensional correlations would induce a modulation of the sawtooth \cite{Fennell:2014fy}. From the magnetic Bragg peaks, we_ found that the moment are ordering in a "up-down" fashion, behaviour of Ising spins, and that both rare earth sites are acting independently. As these sites form zigzag chains along the $c$-axis, the simplest model we_ can use to represent these systems is the 1D ANNNI model \cite{Selke:1988tx}. In the original iteration of the 1D ANNNI model, the spins are collinear in a chain with two interactions: $J_1$ with the nearest neighbour and $J_2$ with the next-nearest neighbours. This model is equivalent to a zigzag chain where $J_1$ is along the diagonal connection and $J_2$ along the $c$-axis.

Diffuse scattering can be harnessed to determine useful information, not only concerning the dimensionality of the interactions, but also the size and direction of the associated moments. Furthermore, it exposes the remaining magnetism from the sites which do not order. 

\begin{figure*}[t] 
    \centering
    \includegraphics{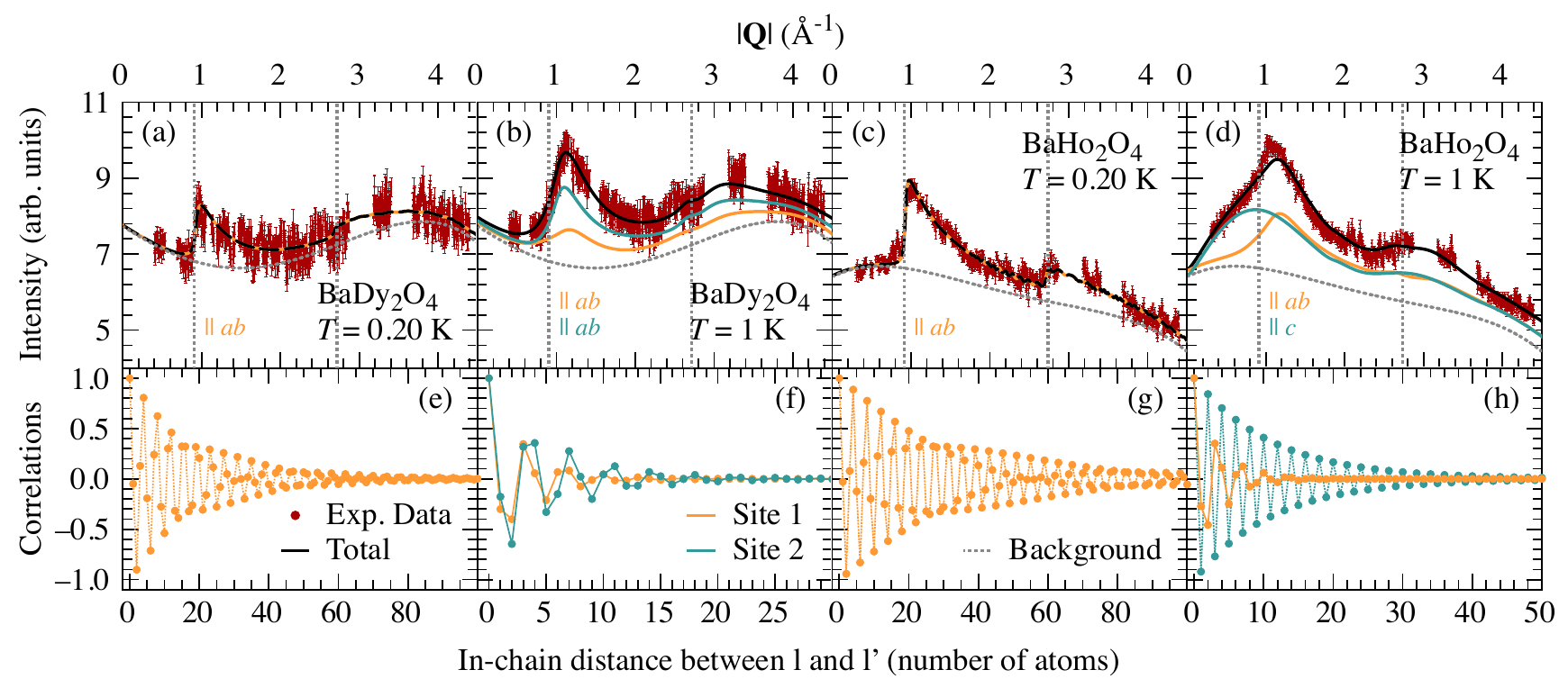} 
    \caption[Selection of diffuse scattering spectra below and above the transition temperatures for \BDO{} and\BHO{}, and their fits with the ANNNI model]{Selection of diffuse scattering spectra below and above the transition temperatures for \BDO{} and \BHO{}, and their fits with the ANNNI model. The red dots are the powder diffraction pattern after subtraction of the nuclear and magnetic Bragg peaks, which then consists of the diffuse magnetic signal and a background. The gaps in the experimental spectra are due to the strong nuclear peaks from the copper sample holder, which were removed. The dotted gray line is the background and the black line the total calculated scattering function, where the contribution from site 1 is shown in orange and from site 2 in blue. Figures (a)--(b) show \BDO{} at $T=0.20$~K, which is below the transition, and at 1~K, which is above. A moment in the $ab$-plane is present on site 1 at $T=0.20$~K, which is kept at the same size at 1~K, where an additional moment in the same plane is added for the site 2. Figures (c)--(d) show \BHO{} at $T=0.20$~K and 1~K, with a moment in the $ab$-plane on site 1 and a moment along $c$ on site 2. Vertical lines are placed at $|\mathbf{Q}|~=~\pi/c$ and $|\mathbf{Q}|~=~3\pi/c$. The lower figures (e)--(h) display the correlation between rare earth atoms within a chain as a function of the number of atoms separating them.}
    \label{fig:Diffuse_Scattering}
\end{figure*}

The shape of the diffuse scattering is similar to what has been observed in \SDO{} and \SHO{}. The sawtooth spectra isolated in \hyperref[fig:Diffuse_Scattering]{Fig.~\ref{fig:Diffuse_Scattering}} do not show the high frequency oscillations expected for the case of two- or three-dimensional correlations \cite{Fennell:2014fy}. This is an indication that the fluctuations are mainly 1D. From our \textsc{multiX} calculations, we found that the moments are Ising-like. As these sites form zigzag chains along the $c$-axis, the simplest model representing these systems is the 1D ANNNI model \cite{Selke:1988tx}. In its original iteration, the spins are collinear in a chain with two interactions: $J_1$ between nearest neighbors and $J_2$ between next-nearest neighbors. This model is equivalent to a zigzag chain where $J_1$ is along the diagonal connection and $J_2$ along the direction of the chain, which in our case is the $c$-axis.

For the analysis of the diffuse scattering, we calculated the partial differential neutron scattering cross-section per solid angle $\Omega$, per unit energy $E$ \cite{Yamani:2010vg}:

\begin{equation}
\frac{\mathrm{d}^2\sigma}{\mathrm{d}\Omega \mathrm{d}E_f} = \frac{|\mathbf{k}_f|}{|\mathbf{k}_i|}\mathrm{e}^{-2W(\mathbf{Q})}\sum_{\alpha\beta}\left(\delta_{\alpha\beta}-\frac{Q_\alpha Q_\beta}{|\mathbf{Q}|^2}\right)S_{\mathrm{mag}}^{\alpha\beta}(\bf{Q},\omega),
\end{equation}
which is summed over the Cartesian coordinates $\alpha$ and $\beta$. Here, the scattering function $S_{\mathrm{mag}}^{\alpha\beta}(\mathbf{Q},\omega)$ is defined as

\begin{multline}
S_{\mathrm{mag}}^{\alpha\beta}(\mathbf{Q},\omega) = \left(\frac{\gamma_n r_0 g}{2}\right)\int \mathrm{d}t \mathrm{e}^{-i\omega t} \\
\times \sum_{ll'}f_{l}^*(\mathbf{Q})f_{l'}^*(\mathbf{Q}) \mathrm{e}^{i\mathbf{Q}\cdot(\mathbf{r}_{l}-\mathbf{r}_{l'})}\langle S_l^\alpha(0) S_{l'}^\beta (t) \rangle,
\end{multline} 
where $\langle S_l^\alpha(0) S_{l'}^\beta (t) \rangle$ is the spin-spin correlation function. Also, $\gamma_n$ is the gyromagnetic ratio of the neutron, $r_0$ the classical electron radius, $g$ the Land\'e $g$-factor, $\mathbf{k}_i$ and $\mathbf{k}_f$ the wave vectors of the incident and scattered neutrons, $\mathbf{Q}~=~\mathbf{k}_i-\mathbf{k}_f$ the momentum transfer, $f(\mathbf{Q})$ the magnetic form factor, and $\mathrm{e}^{-2W(\mathbf{Q})}$ the Debye-Waller factor. The sum has to be taken over all magnetic atoms. In an elastic case, where the energy transfer is negligible, a static approximation reduces the scattering function to

\begin{multline}
S_{\mathrm{mag}}^{\alpha\beta}(\mathbf{Q}) = \left(\frac{\gamma_n r_0 g}{2}\right) \\
\times \sum_{ll'}f_{l}^*(\mathbf{Q})f_{l'}^*(\mathbf{Q}) \mathrm{e}^{i\mathbf{Q}\cdot(\mathbf{r}_{l}-\mathbf{r}_{l'})}\langle S_l^\alpha S_{l'}^\beta \rangle.
\end{multline}
The spin-spin correlation function can be written as

\begin{equation}
\langle S_l^\alpha S_{l'}^\beta \rangle = |\boldsymbol{\mu}|^2 g_{ll'}(J_1,J_2,T),
\label{Eq:spinspincorel}
\end{equation}
where $g_{ll'}$ is the site dependent correlation function between the spins on positions $l$ and $l'$ for given interactions $J_1$ and $J_2$, at a temperature $T$. In the case of the 1D ANNNI model, the correlation function is known in the thermodynamic limit \cite{Stephenson:1970cs}. The powder average has to be calculated for comparison with the powder data. It is defined as the integral of the scattering function over the solid angle \cite{Haraldsen:2005kt}:
\begin{equation}
\bar{S}(|\mathbf{Q}|) = \int\frac{\mathrm{d}\Omega_{\mathbf{\hat{Q}}}}{4\pi} S(\mathbf{Q}). 
\end{equation}

The diffuse scattering is isolated from the diffraction pattern by subtracting the contributions from the nuclear and magnetic Bragg peaks. This results in the diffuse scattering intensity plus a background, as shown in \hyperref[fig:Diffuse_Scattering]{Fig.~\ref{fig:Diffuse_Scattering}}. This figure shows the diffuse scattering for temperatures above and below the formation of order. Strong peaks originating from the copper sample holder were excluded from the spectra. This allows us to fit the residual spectrum using the magnetic interactions $J_1$ and $J_2$, and the moment size $|\boldsymbol{\mu}|$, as adjustable parameters. The background is fitted with a smooth fourth order polynomial.

A closer inspection of the model shows that it is strongly dependent on the interaction ratio $J_2/J_1$ and is fairly insensitive to their individual values. Various combinations of $J_1$ and $J_2$, for a given interaction ratio, will produce a curve with a similar deviation $\chi^2$ between the model and the data. Unfortunately, this means that it is not possible to determine the absolute size of $J_1$ and $J_2$, but only their ratio $J_2/J_1$. However, the calculated spectra are sensitive to the sign of the interaction constants, i.e. whether they are ferromagnetic or antiferromagnetic. Note the sharp features in the diffuse scattering which are located near values of $|\mathbf{Q}|~=~(\mathrm{n}+\frac{1}{2})c^* = (2\mathrm{n}+1)\pi/c$. These positions are indicated by the vertical dashed lines in \hyperref[fig:Diffuse_Scattering]{Fig.~\ref{fig:Diffuse_Scattering}}. For these sawtooth features, the sign of $J_1$ determines if the leading edge is to the left or the right of the dashed line. For a negative $J_1$, the edge is located on the right. That the periodicity is inversely proportional to $c$ is a further indication for the 1D character of these interactions.

Furthermore, the fact that neutrons interact only with magnetic moments which are perpendicular to the momentum transfer strongly affects the profile of the diffuse scattering. The dot product $\mathbf{Q}\cdot(\mathbf{r}_{l}-\mathbf{r}_{l'})$ in $S_{\mathrm{mag}}^{\alpha\beta}(\mathbf{Q},\omega)$ results in a much stronger contribution for any $\mathbf{Q}$ along the rare earth chains, which run parallel to the $c$-axis. A spin orientation along $c$ produces a significantly lower scattering intensity with broader features than a spin in the $ab$-plane. For this reason, a spin along $c$ can be hard to detect in the presence of a large spin component residing in the $ab$-plane. The geometry of the chains also makes it hard to distinguish an alignment of the spins along $a$ or $b$. 

The model contains one zigzag chain. In our structure, the two crystallographically inequivalent rare earth atoms are interacting along separated chains, resulting in a total of four chains within a chemical unit cell --- two on site 1, two on site 2. As the two types of chain are sitting on two inequivalent sites, they are treated with their own set of magnetic interactions $J_1$ and $J_2$, and moment $\boldsymbol{\mu}$. The only difference between the chains A and B of a given rare earth site will be the atomic position, as shown in \hyperref[fig:Structure]{Fig.~\ref{fig:Structure}}. The total calculated spectra will then consist of the sum of the contributions from these four chains, plus a background. The number of atoms contained in a chain is set to such a high number, that the correlations converge to zero over the length of the chain.

While the model includes size and direction of the moments, only the sizes are fitted, with the directions fixed to the easy-axis obtained from the CEF calculations. The contributions from each site are shown as the orange and blue curves in \hyperref[fig:Diffuse_Scattering]{Fig.~\ref{fig:Diffuse_Scattering}}. In order to keep the size of the moments stable, the polynomial background was fitted at a temperature where there is a clear and known contribution from both sites. The background was kept fixed afterwards. The size of the moments in Eq.~\ref{Eq:spinspincorel} was scaled so that they match the values from the \textsc{FullProf} refinements. The results of this procedure are presented in \hyperref[fig:Diffuse_Scattering_Param]{Fig.~\ref{fig:Diffuse_Scattering_Param}}.

Since the systems do not show significant structural change with temperature, the ratio of the exchange constants are therefore expected to remain constant. This ratio was obtained while simultaneously fitting several spectra recorded at a number of temperatures. This strategy was chosen to alleviate problems which could skew this value if too many spectra at temperatures close to the phase transition were chosen. Spectra taken at the lowest measured temperatures are also discarded, as different chains will eventually start to interact with each other.

%%%%%%%%%%%%%%%%%%%%%%%%%%%%%%%%%%%%%%%%%%%%%
\subsubsection{\texorpdfstring{\BDO}{}} %%%%%
%%%%%%%%%%%%%%%%%%%%%%%%%%%%%%%%%%%%%%%%%%%%%

The sharp sawtooth seen at $T$~=~1~K suggests that only magnetism in the $ab$-plane is present, with no contribution from moments parallel to the $c$-axis, as this would result in broader features. This is also coherent with the direction of the moments obtained from \textsc{multiX}. The ANNNI model was then fitted with moments in the $ab$-plane for both sites. The experimental scattering intensity in \hyperref[fig:Diffuse_Scattering]{Fig.~\ref{fig:Diffuse_Scattering}(a)--(b)} strongly decreases with temperature, but stabilizes at $T$~=~0.30~K. It is interesting to note that the diffuse scattering intensity persists down to $T$~=~0.07~K, after the onset of magnetic Bragg peaks. The diffusive intensity at $T$~=~0.07~K still represented roughly half the size of the diffusive scattering intensity at $T$~=~1~K. As expected, the trend of the magnetic scattering intensity is the opposite to the one seen for the magnetic Bragg peaks --- the intensity of the diffuse magnetic scattering decreases with lower temperatures. This seems to indicate that weight from the diffusive scattering is transferred to the magnetic Bragg peaks.

% In the case of \BDO{}, two components along the $ab$-plane were adjusted. As previously mentioned, a close inspection of the diffuse scattering shows a strong signal down to 0.07~K, representing approximately half of the intensity above the magnetic transition, well below the magnetic saturation temperature of 0.40~K of both order seen with the magnetic Bragg peaks. The sharp sawtooth seen at 1~K is sign that only magnetism in the $ab$-plane is present --- no contribution from moment in the $c$-axis --- and its intensity strongly decreases with temperature, stabilizing at 0.30~K. As expected, the trend is exactly the opposite to the one seen with the magnetic intensity of the ordered states --- with lowering temperature, the intensity of the magnetic Bragg peaks is increasing. This confirms that the sharp feature in the diffuse scattering corresponds to the magnetic Bragg peaks.

This leads to two possible scenarios presented in \hyperref[fig:Diffuse_Scattering_Param]{Fig.~\ref{fig:Diffuse_Scattering_Param}(a)--(b)}. In scenario (1), the size of the magnetic moments, which are responsible for the diffuse scattering on both sites, have the same temperature dependence, as seen in the thermal evolution of the magnetic Bragg peaks. However, this would mean that despite the magnetic order manifested through the presence of Bragg peaks, moments of the size of nearly 4$\mu_\mathrm{B}$ and 2$\mu_\mathrm{B}$ would continue to fluctuate down to $T=$~0.07~K on both sites. Or (2), a case where one of the site is not ordering. Here, the intensity of the diffuse scattering associated with site 2 is decreasing to zero for the lowest temperatures. The diffuse scattering seen at low temperature will then be entirely located on the site 1 which will remain in a strongly fluctuating state with a moment size of 5.2(2)$\mu_\mathrm{B}$. This is the scenario shown in \hyperref[fig:Diffuse_Scattering]{Fig.~\ref{fig:Diffuse_Scattering}(a)--(b)}, as the components are more easily separated, and the plot readable. For this case, the moment on site 1 was fitted at low temperatures, where only one contribution would be present, and kept fixed. This prevents that a similar contribution from the other site mixes in the fitting process.

For case (1), shown in \hyperref[fig:Diffuse_Scattering_Param]{Fig.~\ref{fig:Diffuse_Scattering_Param}(a)}, both sites have antiferromagnetic interactions with their nearest neighbors and next-nearest neighbors. The ratios $J_2/J_1$ are 0.59 and 0.60 for the site 1 and 2, respectively, which are both higher than 1/2. This implies that the atoms in the chains of both sites are interacting as a double N\'eel state, consistent with the \textsc{FullProf} analysis. For case (2), shown in \hyperref[fig:Diffuse_Scattering_Param]{Fig.~\ref{fig:Diffuse_Scattering_Param}(b)}, similar values are obtained with $J_2/J_1$ of 0.55 and 0.61 for the site 1 and 2, respectively. As expected, the site that does not order has a ratio $J_2/J_1$ which is closer to the critical value of 1/2.

The correlation function gives the distance over which sites are correlated, as shown in the lower panels of \hyperref[fig:Diffuse_Scattering]{Fig.~\ref{fig:Diffuse_Scattering}}. Temperature and the ratio $J_2/J_1$ are affecting this length --- it decreases when the temperature increases, and when the ratio $J_2/J_1$ is getting closer to 1/2. This is coherent with the observation that the ordered sites have larger ratio $J_2/J_1$ than the fluctuating sites, thus having correlations on a longer range. For case (1), where both sites remain in a fluctuating state, both will have an interaction range of about 20 neighbors at $T=1$~K, which increases to nearly 150 neighbors at 0.20~K. For the case (2), where one site does not order, the correlation range is shorter at $T=1$~K, with 10 neighbors, and it increases to 100 neighbors at 0.20~K.

Given that the two magnetic orders possess two different propagation vectors, it is likely to consider the two orders on separated sites. This was the interpretation resulting from the magnetic Bragg peaks refinement. Case (1) would then be the correct one. A proper interpretation was here not possible solely based on the diffuse scattering analysis.
% Since the moments carried by the two sites of \BDO{} are on the same plane, a proper interpretation is more complicated than the one for \BHO{}. In the later, we argued the unusualness of having the superposition of the two phases on a single site, as the moments would be added on one chain, and subtracted on the other. The same argument can be applied for \BDO{}, and indicates that case (1) is more likely. %Nonetheless, this merely implies that the two long-range orders are one separated sites, and diffuse scattering could be spreads in any proportions on any sites. The \hyperref[fig:Diffuse_Scattering_Param]{Fig.~\ref{fig:Diffuse_Scattering_Param}(a)} represent only one option of diffuse scattering distribution.

%%%%%%%%%%%%%%%%%%%%%%%%%%%%%%%%%%%%%%%%%%%%%%%%%%%%%%%%%
\subsubsection{\texorpdfstring{\BHO}{}} %%%%%
%%%%%%%%%%%%%%%%%%%%%%%%%%%%%%%%%%%%%%%%%%%%%%%%%%%%%%%%%

\begin{figure}
    \centering
    \includegraphics{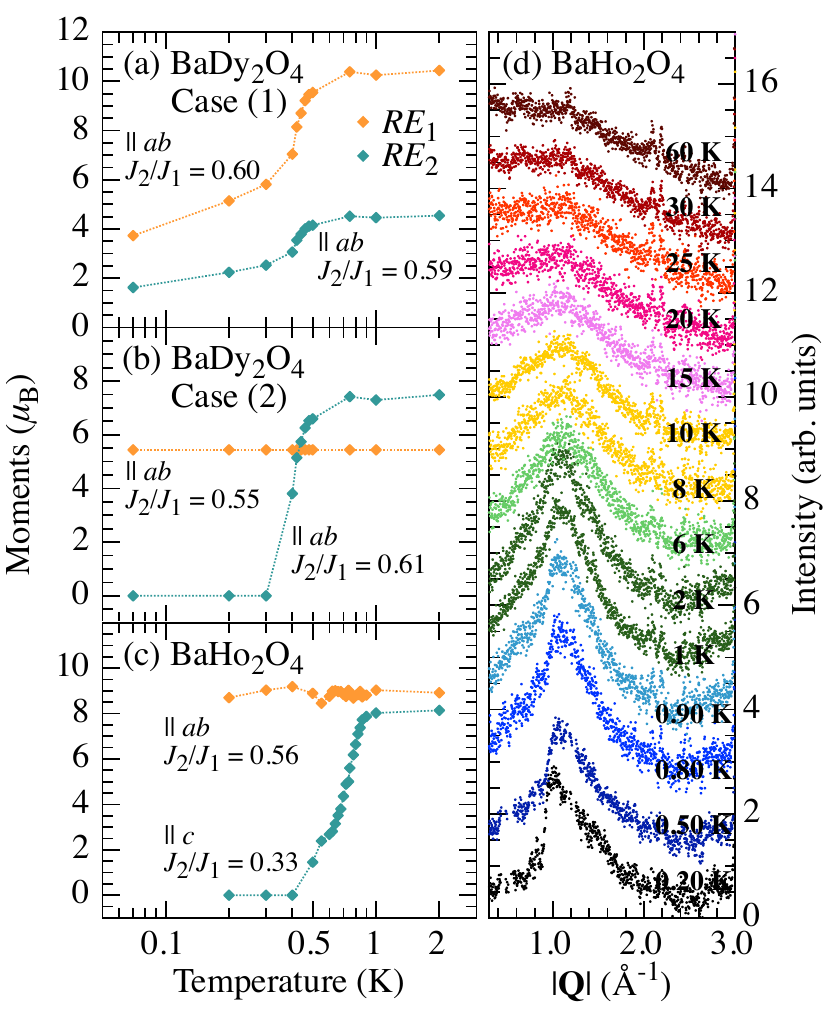} 
    \caption[Parameters from fitting the diffuse magnetic scattering with the 1D ANNNI model]{Parameters from fitting the diffuse magnetic scattering with the 1D ANNNI model. The intensities associated with the different sites are scaled in respect to the total ordered moment in each compound. Figure (a) shows case (1) for \BDO{} where we assumed that both sites order but also remain in a fluctuating state. Figure (b) is the case (2) where the remaining diffuse scattering is assigned to a single site which does not order and has a fixed moment. The other site then sees a strong decrease in the size of the moment with the onset of magnetic order. Figure (c) shows for \BHO{} the decreasing size of the moment of site 2. The moment on site 1 which remains in a strongly fluctuating state is temperature independent. Figure (d) presents a waterfall plot showing the complete evolution of an isolated region of diffuse scattering of \BHO{} from $T=$~0.20~K to 60~K. The correlation are still clearly visible at $T=$~60~K.}
    \label{fig:Diffuse_Scattering_Param}
\end{figure}

For \BHO{}, the analysis of the diffuse scattering at $T=1$~K is shown in \hyperref[fig:Diffuse_Scattering]{Fig.~\ref{fig:Diffuse_Scattering}(d)}. The shape of the spectra indicates the presence of magnetic moments in the $ab$-plane, as well as along $c$-axis. The component along $c$, placed on site 2, decreases in intensity with temperature, as expected due to the magnetic order, which develops with moments along the $c$ direction. The intensity of the component with the moment in the $ab$-plane, placed on site 1, remains constant in the temperature range from 2~K to 0.20~K, as shown in \hyperref[fig:Diffuse_Scattering]{Fig.~\ref{fig:Diffuse_Scattering_Param}(c)}. This result suggests that site 1 does not order down to 0.20~K and remains in a strongly fluctuating state. From our analysis, we can thus conclude that the moment on site 1 lies in the $ab$-plane with a moment size of 8.8(3)$\mu_\mathrm{B}$, which is coherent with the \textsc{multiX} calculations. This is another indication for the fact that the magnetic order of both irreps coexist on site 2. The result of this fit is shown in \hyperref[fig:Diffuse_Scattering_Param]{Fig.~\ref{fig:Diffuse_Scattering_Param}(c)}, where the size of the moment on site 1 remains constant, and the one on site 2 decreases to zero. Here, one can clearly see the different transition temperatures of the two irreps on site 2. The intensity slightly decrease between $T$~=~2~K and 0.80~K, as the order associated with $\mathbf{k}_{\Gamma_2}$ develops, but then sharply drops when the order associated with $\mathbf{k}_{\Gamma_1}$ sets in and the ordered moment becomes much larger. There is no reason to believe that the diffuse scattering intensity on site 2 is not going to zero, but since this signal is broad and of low intensity at low temperature, it could be mistaken as part of the background. Therefore, it cannot be ruled out with certainty that site 2 is still slightly fluctuating.

% \textcolor{TealBlue}{To be rewritten:} The moments sizes established from the fitting parameters were scaled such that the difference between the two plateaux corresponds to the moment size as determined from \textsc{FullProf}. It is interesting to see from the magnetic order intensities and diffuse scattering intensities is that in both case, even in the fully saturated and ordered state seen in Fig.7, a large and stable moment still remains in form of diffuse scattering. Not only it is not ordering at the same pace and st. The remaining moments are of considerable size of 7.4~$\pm$~0.X~$\mu_{\text{B}}$ and 8.4~$\pm$~0.X~$\mu_{\text{B}}$ on the sites of \BDO{} and 8~$\pm$~0.X~$\mu_{\text{B}}$ and 1.5~$\pm$~0.X~$\mu_{\text{B}}$ on \BHO{}. The larger moment on site 1 of \BHO{} is the only component showing no obvious thermal evolution.

Site 1, which does not order, is characterized by all antiferromagnetic interactions with a ratio $J_2/J_1$~=~0.56, similar to \BDO{}'s double N\'eel states. Site 2, which orders, has a much lower ratio of 0.33 --- below the critical value of 1/2 --- again with all antiferromagnetic interactions. It describes a simple N\'eel order, consistent with the \textsc{FullProf} analysis. The site that does not order also has a $J_2/J_1$ which is much closer to the critical value of 1/2. At $T=$~1~K, site 2 has correlations with up to 50 neighbors, while site 1 will interact with only 15 neighbors. At 0.20~K, the correlation length of site 1 increases to more than 150 sites.

The fits were limited to temperatures below 2~K, but for \BHO{}, multiple spectra were recorded at temperatures above 2~K which show strong diffuse scattering and sharp features up to 60~K, as can be seen in \hyperref[fig:Diffuse_Scattering_Param]{Fig.~\ref{fig:Diffuse_Scattering_Param}(d)}. However, the ANNNI model with fixed $J_1$ and $J_2$ is unable to describe the observed spectra. A good fit would require unphysically large magnetic moment sizes --- or a significant change in the values of $J_1$ and $J_2$, which is not expected, as there is no significant change in the crystal structure. On the other hand, it is possible that the higher temperatures are causing the population of higher CEF levels, which from the \textsc{multiX} calculation have been shown to be rather close in energy to the ground state. This could result in a set of more complex interactions and also affect their strength.

Despite the fitting constraint of having fixed interaction constants, the $\chi^2$ only slightly increases from not having a free ratio. Fixing the interaction constants will nevertheless affect the fitting process, as the peak of the sharp feature slightly drifts to higher $|\mathbf{Q}|$ with increasing temperature. While the results are not significantly affected in \BDO{} up to 1~K, this is mainly seen in \BHO{}. The shape of the diffuse scattering peak at 1~K would be more adequately represented, for a fixed $J_1$, with a slightly lower $J_2$ on site 1, and a slightly higher $J_2$ on site 2. The interaction constants would have a slow and linear shift in temperature from 0.20~K to 2~K, with a ratio moving away from 1/2 as temperature increases. While the change is not substantial, it affects the position of the large features significantly and provides a better adjustment to the data, which would be required if the model would be applied at temperatures higher than 2~K. The calculations are also robust while cell parameters and rare earth positions are repositioned within the error bar --- no change in the quality of the fit is noticeable, regardless of the temperature. The apparent shift in $J_2/J_1$ can thus not be explained by structural change.

Nonetheless, the quality of the fits at lower temperature is an indication for the presence of strong 1D correlations in these systems, as was observed in the case of \SDO{} and \SHO{} \cite{Fennell:2014fy}. While additional interactions $J_3$ and $J_4$ between the chains, as defined in \hyperref[fig:Structure]{Fig.~~\ref{fig:Structure}}, are expected to be non-negligible at very low temperatures \cite{Wen:2015wl}, the ANNNI model still adequately accounts for the diffusive part of our spectra.

\section{Discussion and conclusion} %%%%%
%%%%%%%%%%%%%%%%%%%%%%%%%%%%%%%%%%%%%%%%%

In conclusion, we have performed CEF calculations using \textsc{multiX} to interpret our inelastic neutron scattering spectra. From these, we find a site-dependent magnetic anisotropy as well as the presence of an easy-axis, which is confined to either the $ab$-plane, or along the $c$-axis. We measured the temperature dependence of the neutron powder diffraction signal which we used to perform a refinement of the magnetic Bragg peaks. These spectra show strong diffusive components, which we fitted using a 1D ANNNI model, confirming the coexistence of long-range and short-range magnetic orders in both samples. In \BDO{}, we observe magnetic order with two different wave vectors but a single transition temperature of $T_\mathrm{N}$~=~0.48~K. The ordered moments are restricted to the $ab$-plane. Magnetic fluctuations remain important to the lowest temperatures, even after the ordered moments saturate. The direction of the fluctuating moment is shown to be in the $ab$-plane, the same plane as the long-range order. For \BHO{}, the magnetic order has to two different irreps with the same propagation vector $\mathbf{k}_0$~=~(0,0,0). The transitions occur at different temperatures, one at $T_\mathrm{N\Gamma_1}$~=~0.84~K, and the other at $T_\mathrm{N\Gamma_1}\sim1.3$~K. For both, the moments order along the $c$-axis. Magnetic fluctuations also remain after the order saturates, predominantly in the $ab$-plane. Agreements between the intensities of the modeled diffuse scattering and of the Bragg peaks indicate that interactions in these samples are for the most part 1D and confined within a specific zigzag chain.

This shares similarities with the behavior seen in \SDHO{}. In \SHO{}, the sites 1 and 2 are host of a simple N\'eel long-range order along the $c$-axis, and a double N\'eel short-range order in the $ab$-plane, respectively \cite{Wen:2015wl}. On the other hand, no long-range order has been found in \SDO{}, but a detailed analysis of the diffuse scattering shows the presence of short-range correlations on both sites, with a spin alignment nonetheless in the same directions as the sites of \SHO \cite{Gauthier:2017vz}. 

The compounds SrHo$_2$O$_4$ \cite{Ghosh:2011ew,Young:2012ji,Young:2013hw,Fennell:2014fy,Wen:2015wl}, SrEr$_2$O$_4$ \cite{Petrenko2008p612,Hayes:2011gf}, and BaNd$_2$O$_4$ \cite{Aczel:2014ko}, either through powders or single crystals neutron diffraction, are showing the coexistence of long-range (or quasi long-range) order and short-range order. So far, however, SrYb$_2$O$_4$ \cite{QuinteroCastro:2012dd} is the only system where two long-range orders were reported. Other compounds such as BaTb$_2$O$_4$ \cite{Aczel:2015ko} and SrDy$_2$O$_4$ \cite{Fennell:2014fy} are only exhibiting short-range order, while BaTm$_2$O$_4$ \cite{Li:2015dh} does not show any sign of ordering. Of all the compounds studied in this family, only SrGd$_2$O$_4$ shows two separated magnetic transition temperatures in zero-field, as seen from specific heat measurements \cite{Young:2014bc}. Its magnetic structure still remains unsolved. Another compound which stands out is SrTb$_2$O$_4$ \cite{Li:2014gs} where the long-range order is incommensurate.

With our results on \BDO{} and \BHO{}, we confirm the presence of two different types of long-range magnetic order --- with different transition temperatures in one case --- coexisting with fluctuating moments down to the lowest measured temperatures, unveiling two compounds with interesting and complex magnetic interactions.
% On the experimental front, the cryogenic constraints prevent us to follow the behavior of the magnetic correlations to lower temperatures, where it is unknown if the magnetic fluctuations would eventually order or if the spins would freeze, unordered, as we approach to the limit of 0~K.
Measurement on single crystals will soon follow. This will allow us to describe more precisely the evolution of the diffuse scattering, especially with the use of polarized neutron scattering measurements which will help to separate the contributions from the different sites. This would also allow us to study the formation of magnetic domains which were seen in \SDHO{} \cite{Wen:2015wl,Gauthier:2017vz}.

A conservative scenario for both samples is that only one of the sites orders --- hosting two types of magnetic order --- and the other one remains fluctuating down to the lowest measured temperatures. However, from a combination of CEF calculations, refinement of the Bragg peaks, and a description of the nature of the diffuse scattering, there are strong indications for more exotic magnetic behaviors. In the case of \BDO{}, we suggest the interesting possibility that both sites order, but keep fluctuating down to the lowest temperatures. This could then be a case of magnetic fragmentation, which as been observed in the spin liquid phase of a pyrochlore, where magnetic order and fluctuations coexist on the same site \cite{BrooksBartlett:2014kf,Petit:2016dg}. Indications for a \textit{classical} spin liquid ground state was indeed found from an investigation by ultrasound velocity measurements in the related compound \SDO{} \cite{Bidaud:2016ki}, as well as by $\mu$SR \cite{Gauthier:2017uh}. \BDO{} might presents a new class of magnetic ground state whose nature is yet to be determined, which surely present an interesting challenge.

Concerning \BHO{}, we propose a scenario in which only one of the two magnetic sites is ordering, initially at $T_\mathrm{N\Gamma_2}\sim1.3$~K. A second order on the same site, with a different irrep, takes place at $T_\mathrm{N\Gamma_1}~=~0.84$~K, leading to a separation of the magnetic phase. The two phase fractions evolve until they reach proportions of 20\%--80\%, where the fraction associated with $\Gamma_1$ becomes dominant. This scenario might be the result of the growing influence of the inter-chain interactions as the temperature is lowered. These interactions are required for long-range order, and their evolving influence could suddenly privilege one order over the other.

%No entropy, no 2nd transition, coherent with fluctuation.

\begin{acknowledgments}
The research at the Universit\'e de Montr\'eal received support from the Natural Sciences and Engineering Research Council of Canada (NSERC), the Fonds de recherche du Qu\'ebec – Nature et technologies (FRQNT), and the Canada Research Chair Foundation, and the work at PSI received support from the Swiss National Foundation (SNF Grant No. 138018).
\end{acknowledgments}

\bibliographystyle{apsrev4-1.bst}
\bibliography{Bibliography.bib}

\begin{thebibliography}{53}%
\makeatletter
\providecommand \@ifxundefined [1]{%
 \@ifx{#1\undefined}
}%
\providecommand \@ifnum [1]{%
 \ifnum #1\expandafter \@firstoftwo
 \else \expandafter \@secondoftwo
 \fi
}%
\providecommand \@ifx [1]{%
 \ifx #1\expandafter \@firstoftwo
 \else \expandafter \@secondoftwo
 \fi
}%
\providecommand \natexlab [1]{#1}%
\providecommand \enquote  [1]{``#1''}%
\providecommand \bibnamefont  [1]{#1}%
\providecommand \bibfnamefont [1]{#1}%
\providecommand \citenamefont [1]{#1}%
\providecommand \href@noop [0]{\@secondoftwo}%
\providecommand \href [0]{\begingroup \@sanitize@url \@href}%
\providecommand \@href[1]{\@@startlink{#1}\@@href}%
\providecommand \@@href[1]{\endgroup#1\@@endlink}%
\providecommand \@sanitize@url [0]{\catcode `\\12\catcode `\$12\catcode
  `\&12\catcode `\#12\catcode `\^12\catcode `\_12\catcode `\%12\relax}%
\providecommand \@@startlink[1]{}%
\providecommand \@@endlink[0]{}%
\providecommand \url  [0]{\begingroup\@sanitize@url \@url }%
\providecommand \@url [1]{\endgroup\@href {#1}{\urlprefix }}%
\providecommand \urlprefix  [0]{URL }%
\providecommand \Eprint [0]{\href }%
\providecommand \doibase [0]{http://dx.doi.org/}%
\providecommand \selectlanguage [0]{\@gobble}%
\providecommand \bibinfo  [0]{\@secondoftwo}%
\providecommand \bibfield  [0]{\@secondoftwo}%
\providecommand \translation [1]{[#1]}%
\providecommand \BibitemOpen [0]{}%
\providecommand \bibitemStop [0]{}%
\providecommand \bibitemNoStop [0]{.\EOS\space}%
\providecommand \EOS [0]{\spacefactor3000\relax}%
\providecommand \BibitemShut  [1]{\csname bibitem#1\endcsname}%
\let\auto@bib@innerbib\@empty
%</preamble>
\bibitem [{\citenamefont {Karunadasa}\ \emph {et~al.}(2005)\citenamefont
  {Karunadasa}, \citenamefont {Huang}, \citenamefont {Ueland}, \citenamefont
  {Lynn}, \citenamefont {Schiffer}, \citenamefont {Regan},\ and\ \citenamefont
  {Cava}}]{Karunadasa2005p41}%
  \BibitemOpen
  \bibfield  {author} {\bibinfo {author} {\bibfnamefont {H.}~\bibnamefont
  {Karunadasa}}, \bibinfo {author} {\bibfnamefont {Q.}~\bibnamefont {Huang}},
  \bibinfo {author} {\bibfnamefont {B.~G.}\ \bibnamefont {Ueland}}, \bibinfo
  {author} {\bibfnamefont {J.~W.}\ \bibnamefont {Lynn}}, \bibinfo {author}
  {\bibfnamefont {P.}~\bibnamefont {Schiffer}}, \bibinfo {author}
  {\bibfnamefont {K.~A.}\ \bibnamefont {Regan}}, \ and\ \bibinfo {author}
  {\bibfnamefont {R.~J.}\ \bibnamefont {Cava}},\ }\href@noop {} {\bibfield
  {journal} {\bibinfo  {journal} {Physical Review B}\ }\textbf {\bibinfo
  {volume} {71}},\ \bibinfo {pages} {144414} (\bibinfo {year}
  {2005})}\BibitemShut {NoStop}%
\bibitem [{\citenamefont {Petrenko}\ \emph {et~al.}(2008)\citenamefont
  {Petrenko}, \citenamefont {Balakrishnan}, \citenamefont {Wilson},
  \citenamefont {de~Brion}, \citenamefont {Suard},\ and\ \citenamefont
  {Chapon}}]{Petrenko2008p612}%
  \BibitemOpen
  \bibfield  {author} {\bibinfo {author} {\bibfnamefont {O.~A.}\ \bibnamefont
  {Petrenko}}, \bibinfo {author} {\bibfnamefont {G.}~\bibnamefont
  {Balakrishnan}}, \bibinfo {author} {\bibfnamefont {N.~R.}\ \bibnamefont
  {Wilson}}, \bibinfo {author} {\bibfnamefont {S.}~\bibnamefont {de~Brion}},
  \bibinfo {author} {\bibfnamefont {E.}~\bibnamefont {Suard}}, \ and\ \bibinfo
  {author} {\bibfnamefont {L.~C.}\ \bibnamefont {Chapon}},\ }\href@noop {}
  {\bibfield  {journal} {\bibinfo  {journal} {Physical Review B}\ }\textbf
  {\bibinfo {volume} {78}},\ \bibinfo {pages} {184410} (\bibinfo {year}
  {2008})}\BibitemShut {NoStop}%
\bibitem [{\citenamefont {Ghosh}\ \emph {et~al.}(2011)\citenamefont {Ghosh},
  \citenamefont {Zhou}, \citenamefont {Balicas}, \citenamefont {Hill},
  \citenamefont {Gardner}, \citenamefont {Qiu},\ and\ \citenamefont
  {Wiebe}}]{Ghosh:2011ew}%
  \BibitemOpen
  \bibfield  {author} {\bibinfo {author} {\bibfnamefont {S.}~\bibnamefont
  {Ghosh}}, \bibinfo {author} {\bibfnamefont {H.~D.}\ \bibnamefont {Zhou}},
  \bibinfo {author} {\bibfnamefont {L.}~\bibnamefont {Balicas}}, \bibinfo
  {author} {\bibfnamefont {S.}~\bibnamefont {Hill}}, \bibinfo {author}
  {\bibfnamefont {J.~S.}\ \bibnamefont {Gardner}}, \bibinfo {author}
  {\bibfnamefont {Y.}~\bibnamefont {Qiu}}, \ and\ \bibinfo {author}
  {\bibfnamefont {C.~R.}\ \bibnamefont {Wiebe}},\ }\href@noop {} {\bibfield
  {journal} {\bibinfo  {journal} {Journal of Physics: Condensed Matter}\
  }\textbf {\bibinfo {volume} {23}},\ \bibinfo {pages} {164203} (\bibinfo
  {year} {2011})}\BibitemShut {NoStop}%
\bibitem [{\citenamefont {Hayes}\ \emph {et~al.}(2011)\citenamefont {Hayes},
  \citenamefont {Balakrishnan}, \citenamefont {Deen}, \citenamefont {Manuel},
  \citenamefont {Chapon},\ and\ \citenamefont {Petrenko}}]{Hayes:2011gf}%
  \BibitemOpen
  \bibfield  {author} {\bibinfo {author} {\bibfnamefont {T.~J.}\ \bibnamefont
  {Hayes}}, \bibinfo {author} {\bibfnamefont {G.}~\bibnamefont {Balakrishnan}},
  \bibinfo {author} {\bibfnamefont {P.~P.}\ \bibnamefont {Deen}}, \bibinfo
  {author} {\bibfnamefont {P.}~\bibnamefont {Manuel}}, \bibinfo {author}
  {\bibfnamefont {L.~C.}\ \bibnamefont {Chapon}}, \ and\ \bibinfo {author}
  {\bibfnamefont {O.~A.}\ \bibnamefont {Petrenko}},\ }\href@noop {} {\bibfield
  {journal} {\bibinfo  {journal} {Physical Review B}\ }\textbf {\bibinfo
  {volume} {84}},\ \bibinfo {pages} {174435} (\bibinfo {year}
  {2011})}\BibitemShut {NoStop}%
\bibitem [{\citenamefont {Young}\ \emph {et~al.}(2012)\citenamefont {Young},
  \citenamefont {Chapon},\ and\ \citenamefont {Petrenko}}]{Young:2012ji}%
  \BibitemOpen
  \bibfield  {author} {\bibinfo {author} {\bibfnamefont {O.}~\bibnamefont
  {Young}}, \bibinfo {author} {\bibfnamefont {L.~C.}\ \bibnamefont {Chapon}}, \
  and\ \bibinfo {author} {\bibfnamefont {O.~A.}\ \bibnamefont {Petrenko}},\
  }\href@noop {} {\bibfield  {journal} {\bibinfo  {journal} {Journal of
  Physics: Conference Series}\ }\textbf {\bibinfo {volume} {391}},\ \bibinfo
  {pages} {012081} (\bibinfo {year} {2012})}\BibitemShut {NoStop}%
\bibitem [{\citenamefont {Young}\ \emph {et~al.}(2013)\citenamefont {Young},
  \citenamefont {Wildes}, \citenamefont {Manuel}, \citenamefont {Ouladdiaf},
  \citenamefont {Khalyavin}, \citenamefont {Balakrishnan},\ and\ \citenamefont
  {Petrenko}}]{Young:2013hw}%
  \BibitemOpen
  \bibfield  {author} {\bibinfo {author} {\bibfnamefont {O.}~\bibnamefont
  {Young}}, \bibinfo {author} {\bibfnamefont {A.~R.}\ \bibnamefont {Wildes}},
  \bibinfo {author} {\bibfnamefont {P.}~\bibnamefont {Manuel}}, \bibinfo
  {author} {\bibfnamefont {B.}~\bibnamefont {Ouladdiaf}}, \bibinfo {author}
  {\bibfnamefont {D.~D.}\ \bibnamefont {Khalyavin}}, \bibinfo {author}
  {\bibfnamefont {G.}~\bibnamefont {Balakrishnan}}, \ and\ \bibinfo {author}
  {\bibfnamefont {O.~A.}\ \bibnamefont {Petrenko}},\ }\href@noop {} {\bibfield
  {journal} {\bibinfo  {journal} {Physical Review B}\ }\textbf {\bibinfo
  {volume} {88}},\ \bibinfo {pages} {024411} (\bibinfo {year}
  {2013})}\BibitemShut {NoStop}%
\bibitem [{\citenamefont {Wen}\ \emph {et~al.}(2015)\citenamefont {Wen},
  \citenamefont {Tian}, \citenamefont {Garlea}, \citenamefont {Koohpayeh},
  \citenamefont {McQueen}, \citenamefont {Li}, \citenamefont {Yan},
  \citenamefont {Rodriguez-Rivera}, \citenamefont {Vaknin},\ and\ \citenamefont
  {Broholm}}]{Wen:2015wl}%
  \BibitemOpen
  \bibfield  {author} {\bibinfo {author} {\bibfnamefont {J.~J.}\ \bibnamefont
  {Wen}}, \bibinfo {author} {\bibfnamefont {W.}~\bibnamefont {Tian}}, \bibinfo
  {author} {\bibfnamefont {V.~O.}\ \bibnamefont {Garlea}}, \bibinfo {author}
  {\bibfnamefont {S.~M.}\ \bibnamefont {Koohpayeh}}, \bibinfo {author}
  {\bibfnamefont {T.~M.}\ \bibnamefont {McQueen}}, \bibinfo {author}
  {\bibfnamefont {H.-F.}\ \bibnamefont {Li}}, \bibinfo {author} {\bibfnamefont
  {J.~Q.}\ \bibnamefont {Yan}}, \bibinfo {author} {\bibfnamefont {J.~A.}\
  \bibnamefont {Rodriguez-Rivera}}, \bibinfo {author} {\bibfnamefont
  {D.}~\bibnamefont {Vaknin}}, \ and\ \bibinfo {author} {\bibfnamefont {C.~L.}\
  \bibnamefont {Broholm}},\ }\href@noop {} {\bibfield  {journal} {\bibinfo
  {journal} {Physical Review B}\ }\textbf {\bibinfo {volume} {91}},\ \bibinfo
  {pages} {054424} (\bibinfo {year} {2015})}\BibitemShut {NoStop}%
\bibitem [{\citenamefont {Fennell}\ \emph
  {et~al.}(2014{\natexlab{a}})\citenamefont {Fennell}, \citenamefont
  {Pomjakushin}, \citenamefont {Uldry}, \citenamefont {Delley}, \citenamefont
  {Pr{\'e}vost}, \citenamefont {D{\'e}silets-Benoit}, \citenamefont {Bianchi},
  \citenamefont {Bewley}, \citenamefont {Hansen}, \citenamefont {Klimczuk},
  \citenamefont {Cava},\ and\ \citenamefont {Kenzelmann}}]{Fennell:2014fy}%
  \BibitemOpen
  \bibfield  {author} {\bibinfo {author} {\bibfnamefont {A.}~\bibnamefont
  {Fennell}}, \bibinfo {author} {\bibfnamefont {V.~Y.}\ \bibnamefont
  {Pomjakushin}}, \bibinfo {author} {\bibfnamefont {A.}~\bibnamefont {Uldry}},
  \bibinfo {author} {\bibfnamefont {B.}~\bibnamefont {Delley}}, \bibinfo
  {author} {\bibfnamefont {B.}~\bibnamefont {Pr{\'e}vost}}, \bibinfo {author}
  {\bibfnamefont {A.}~\bibnamefont {D{\'e}silets-Benoit}}, \bibinfo {author}
  {\bibfnamefont {A.~D.}\ \bibnamefont {Bianchi}}, \bibinfo {author}
  {\bibfnamefont {R.~I.}\ \bibnamefont {Bewley}}, \bibinfo {author}
  {\bibfnamefont {B.~R.}\ \bibnamefont {Hansen}}, \bibinfo {author}
  {\bibfnamefont {T.}~\bibnamefont {Klimczuk}}, \bibinfo {author}
  {\bibfnamefont {R.~J.}\ \bibnamefont {Cava}}, \ and\ \bibinfo {author}
  {\bibfnamefont {M.}~\bibnamefont {Kenzelmann}},\ }\href@noop {} {\bibfield
  {journal} {\bibinfo  {journal} {Physical Review B}\ }\textbf {\bibinfo
  {volume} {89}},\ \bibinfo {pages} {224511} (\bibinfo {year}
  {2014}{\natexlab{a}})}\BibitemShut {NoStop}%
\bibitem [{\citenamefont {Li}\ \emph {et~al.}(2015)\citenamefont {Li},
  \citenamefont {Senyshyn}, \citenamefont {Fabelo}, \citenamefont {Persson},
  \citenamefont {Hou}, \citenamefont {Boehm}, \citenamefont {Schmalzl},
  \citenamefont {Schmidt}, \citenamefont {Vassalli}, \citenamefont {Thakuria},
  \citenamefont {Sun}, \citenamefont {Wang}, \citenamefont {Khazaradze},
  \citenamefont {Schmitz}, \citenamefont {Zhang}, \citenamefont {Roth},
  \citenamefont {Roca},\ and\ \citenamefont {Wildes}}]{Li:2015dh}%
  \BibitemOpen
  \bibfield  {author} {\bibinfo {author} {\bibfnamefont {H.-F.}\ \bibnamefont
  {Li}}, \bibinfo {author} {\bibfnamefont {A.}~\bibnamefont {Senyshyn}},
  \bibinfo {author} {\bibfnamefont {O.}~\bibnamefont {Fabelo}}, \bibinfo
  {author} {\bibfnamefont {J.}~\bibnamefont {Persson}}, \bibinfo {author}
  {\bibfnamefont {B.}~\bibnamefont {Hou}}, \bibinfo {author} {\bibfnamefont
  {M.}~\bibnamefont {Boehm}}, \bibinfo {author} {\bibfnamefont
  {K.}~\bibnamefont {Schmalzl}}, \bibinfo {author} {\bibfnamefont
  {W.}~\bibnamefont {Schmidt}}, \bibinfo {author} {\bibfnamefont {J.-P.}\
  \bibnamefont {Vassalli}}, \bibinfo {author} {\bibfnamefont {P.}~\bibnamefont
  {Thakuria}}, \bibinfo {author} {\bibfnamefont {X.}~\bibnamefont {Sun}},
  \bibinfo {author} {\bibfnamefont {L.}~\bibnamefont {Wang}}, \bibinfo {author}
  {\bibfnamefont {G.}~\bibnamefont {Khazaradze}}, \bibinfo {author}
  {\bibfnamefont {B.}~\bibnamefont {Schmitz}}, \bibinfo {author} {\bibfnamefont
  {C.}~\bibnamefont {Zhang}}, \bibinfo {author} {\bibfnamefont
  {G.}~\bibnamefont {Roth}}, \bibinfo {author} {\bibfnamefont {J.~G.}\
  \bibnamefont {Roca}}, \ and\ \bibinfo {author} {\bibfnamefont
  {A.}~\bibnamefont {Wildes}},\ }\href@noop {} {\bibfield  {journal} {\bibinfo
  {journal} {J. Mater. Chem. C}\ }\textbf {\bibinfo {volume} {3}},\ \bibinfo
  {pages} {7658} (\bibinfo {year} {2015})}\BibitemShut {NoStop}%
\bibitem [{\citenamefont {Gauthier}\ \emph
  {et~al.}(2017{\natexlab{a}})\citenamefont {Gauthier}, \citenamefont
  {Fennell}, \citenamefont {Pr{\'e}vost}, \citenamefont {Uldry}, \citenamefont
  {Delley}, \citenamefont {Sibille}, \citenamefont {D{\'e}silets-Benoit},
  \citenamefont {Dabkowska}, \citenamefont {Nilsen}, \citenamefont {Regnault},
  \citenamefont {White}, \citenamefont {Niedermayer}, \citenamefont
  {Pomjakushin}, \citenamefont {Bianchi},\ and\ \citenamefont
  {Kenzelmann}}]{Gauthier:2017vz}%
  \BibitemOpen
  \bibfield  {author} {\bibinfo {author} {\bibfnamefont {N.}~\bibnamefont
  {Gauthier}}, \bibinfo {author} {\bibfnamefont {A.}~\bibnamefont {Fennell}},
  \bibinfo {author} {\bibfnamefont {B.}~\bibnamefont {Pr{\'e}vost}}, \bibinfo
  {author} {\bibfnamefont {A.~C.}\ \bibnamefont {Uldry}}, \bibinfo {author}
  {\bibfnamefont {B.}~\bibnamefont {Delley}}, \bibinfo {author} {\bibfnamefont
  {R.}~\bibnamefont {Sibille}}, \bibinfo {author} {\bibfnamefont
  {A.}~\bibnamefont {D{\'e}silets-Benoit}}, \bibinfo {author} {\bibfnamefont
  {H.~A.}\ \bibnamefont {Dabkowska}}, \bibinfo {author} {\bibfnamefont {G.~J.}\
  \bibnamefont {Nilsen}}, \bibinfo {author} {\bibfnamefont {L.-P.}\
  \bibnamefont {Regnault}}, \bibinfo {author} {\bibfnamefont {J.~S.}\
  \bibnamefont {White}}, \bibinfo {author} {\bibfnamefont {C.}~\bibnamefont
  {Niedermayer}}, \bibinfo {author} {\bibfnamefont {V.}~\bibnamefont
  {Pomjakushin}}, \bibinfo {author} {\bibfnamefont {A.~D.}\ \bibnamefont
  {Bianchi}}, \ and\ \bibinfo {author} {\bibfnamefont {M.}~\bibnamefont
  {Kenzelmann}},\ }\href@noop {} {\bibfield  {journal} {\bibinfo  {journal}
  {Physical Review B}\ }\textbf {\bibinfo {volume} {95}},\ \bibinfo {pages}
  {134430} (\bibinfo {year} {2017}{\natexlab{a}})}\BibitemShut {NoStop}%
\bibitem [{\citenamefont {Hayes}\ \emph {et~al.}(2012)\citenamefont {Hayes},
  \citenamefont {Young}, \citenamefont {Balakrishnan},\ and\ \citenamefont
  {Petrenko}}]{Hayes:2012kt}%
  \BibitemOpen
  \bibfield  {author} {\bibinfo {author} {\bibfnamefont {T.~J.}\ \bibnamefont
  {Hayes}}, \bibinfo {author} {\bibfnamefont {O.}~\bibnamefont {Young}},
  \bibinfo {author} {\bibfnamefont {G.}~\bibnamefont {Balakrishnan}}, \ and\
  \bibinfo {author} {\bibfnamefont {O.~A.}\ \bibnamefont {Petrenko}},\
  }\href@noop {} {\bibfield  {journal} {\bibinfo  {journal} {Journal Of The
  Physical Society Of Japan}\ }\textbf {\bibinfo {volume} {81}},\ \bibinfo
  {pages} {024708} (\bibinfo {year} {2012})}\BibitemShut {NoStop}%
\bibitem [{\citenamefont {Quintero-Castro}\ \emph {et~al.}(2012)\citenamefont
  {Quintero-Castro}, \citenamefont {Lake}, \citenamefont {Reehuis},
  \citenamefont {Niazi}, \citenamefont {Ryll}, \citenamefont {Islam},
  \citenamefont {Fennell}, \citenamefont {Kimber}, \citenamefont {Klemke},
  \citenamefont {Ollivier}, \citenamefont {Garcia~Sakai}, \citenamefont
  {Deen},\ and\ \citenamefont {Mutka}}]{QuinteroCastro:2012dd}%
  \BibitemOpen
  \bibfield  {author} {\bibinfo {author} {\bibfnamefont {D.~L.}\ \bibnamefont
  {Quintero-Castro}}, \bibinfo {author} {\bibfnamefont {B.}~\bibnamefont
  {Lake}}, \bibinfo {author} {\bibfnamefont {M.}~\bibnamefont {Reehuis}},
  \bibinfo {author} {\bibfnamefont {A.}~\bibnamefont {Niazi}}, \bibinfo
  {author} {\bibfnamefont {H.}~\bibnamefont {Ryll}}, \bibinfo {author}
  {\bibfnamefont {A.~T. M.~N.}\ \bibnamefont {Islam}}, \bibinfo {author}
  {\bibfnamefont {T.}~\bibnamefont {Fennell}}, \bibinfo {author} {\bibfnamefont
  {S.~A.~J.}\ \bibnamefont {Kimber}}, \bibinfo {author} {\bibfnamefont
  {B.}~\bibnamefont {Klemke}}, \bibinfo {author} {\bibfnamefont
  {J.}~\bibnamefont {Ollivier}}, \bibinfo {author} {\bibfnamefont
  {V.}~\bibnamefont {Garcia~Sakai}}, \bibinfo {author} {\bibfnamefont {P.~P.}\
  \bibnamefont {Deen}}, \ and\ \bibinfo {author} {\bibfnamefont
  {H.}~\bibnamefont {Mutka}},\ }\href@noop {} {\bibfield  {journal} {\bibinfo
  {journal} {Physical Review B}\ }\textbf {\bibinfo {volume} {86}},\ \bibinfo
  {pages} {064203} (\bibinfo {year} {2012})}\BibitemShut {NoStop}%
\bibitem [{\citenamefont {Cheffings}\ \emph {et~al.}(2013)\citenamefont
  {Cheffings}, \citenamefont {Lees}, \citenamefont {Balakrishnan},\ and\
  \citenamefont {Petrenko}}]{Cheffings:2013bw}%
  \BibitemOpen
  \bibfield  {author} {\bibinfo {author} {\bibfnamefont {T.~H.}\ \bibnamefont
  {Cheffings}}, \bibinfo {author} {\bibfnamefont {M.~R.}\ \bibnamefont {Lees}},
  \bibinfo {author} {\bibfnamefont {G.}~\bibnamefont {Balakrishnan}}, \ and\
  \bibinfo {author} {\bibfnamefont {O.~A.}\ \bibnamefont {Petrenko}},\
  }\href@noop {} {\bibfield  {journal} {\bibinfo  {journal} {Journal of
  Physics: Condensed Matter}\ }\textbf {\bibinfo {volume} {25}},\ \bibinfo
  {pages} {256001} (\bibinfo {year} {2013})}\BibitemShut {NoStop}%
\bibitem [{\citenamefont {Bidaud}\ \emph {et~al.}(2016)\citenamefont {Bidaud},
  \citenamefont {Simard}, \citenamefont {Quirion}, \citenamefont {Pr{\'e}vost},
  \citenamefont {Daneau}, \citenamefont {Bianchi}, \citenamefont {Dabkowska},\
  and\ \citenamefont {Quilliam}}]{Bidaud:2016ki}%
  \BibitemOpen
  \bibfield  {author} {\bibinfo {author} {\bibfnamefont {C.}~\bibnamefont
  {Bidaud}}, \bibinfo {author} {\bibfnamefont {O.}~\bibnamefont {Simard}},
  \bibinfo {author} {\bibfnamefont {G.}~\bibnamefont {Quirion}}, \bibinfo
  {author} {\bibfnamefont {B.}~\bibnamefont {Pr{\'e}vost}}, \bibinfo {author}
  {\bibfnamefont {S.}~\bibnamefont {Daneau}}, \bibinfo {author} {\bibfnamefont
  {A.~D.}\ \bibnamefont {Bianchi}}, \bibinfo {author} {\bibfnamefont {H.~A.}\
  \bibnamefont {Dabkowska}}, \ and\ \bibinfo {author} {\bibfnamefont {J.~A.}\
  \bibnamefont {Quilliam}},\ }\href@noop {} {\bibfield  {journal} {\bibinfo
  {journal} {Physical Review B}\ }\textbf {\bibinfo {volume} {93}},\ \bibinfo
  {pages} {060404} (\bibinfo {year} {2016})}\BibitemShut {NoStop}%
\bibitem [{\citenamefont {Petrenko}\ \emph {et~al.}(2017)\citenamefont
  {Petrenko}, \citenamefont {Young}, \citenamefont {Brunt}, \citenamefont
  {Balakrishnan}, \citenamefont {Manuel}, \citenamefont {Khalyavin},\ and\
  \citenamefont {Ritter}}]{Petrenko:2017em}%
  \BibitemOpen
  \bibfield  {author} {\bibinfo {author} {\bibfnamefont {O.~A.}\ \bibnamefont
  {Petrenko}}, \bibinfo {author} {\bibfnamefont {O.}~\bibnamefont {Young}},
  \bibinfo {author} {\bibfnamefont {D.}~\bibnamefont {Brunt}}, \bibinfo
  {author} {\bibfnamefont {G.}~\bibnamefont {Balakrishnan}}, \bibinfo {author}
  {\bibfnamefont {P.}~\bibnamefont {Manuel}}, \bibinfo {author} {\bibfnamefont
  {D.~D.}\ \bibnamefont {Khalyavin}}, \ and\ \bibinfo {author} {\bibfnamefont
  {C.}~\bibnamefont {Ritter}},\ }\href@noop {} {\bibfield  {journal} {\bibinfo
  {journal} {Physical Review B}\ }\textbf {\bibinfo {volume} {95}},\ \bibinfo
  {pages} {104442} (\bibinfo {year} {2017})}\BibitemShut {NoStop}%
\bibitem [{\citenamefont {Gauthier}\ \emph
  {et~al.}(2017{\natexlab{b}})\citenamefont {Gauthier}, \citenamefont
  {Fennell}, \citenamefont {Pr{\'e}vost}, \citenamefont {D{\'e}silets-Benoit},
  \citenamefont {Dabkowska}, \citenamefont {Zaharko}, \citenamefont {Frontzek},
  \citenamefont {Sibille}, \citenamefont {Bianchi},\ and\ \citenamefont
  {Kenzelmann}}]{Gauthier:2017ts}%
  \BibitemOpen
  \bibfield  {author} {\bibinfo {author} {\bibfnamefont {N.}~\bibnamefont
  {Gauthier}}, \bibinfo {author} {\bibfnamefont {A.}~\bibnamefont {Fennell}},
  \bibinfo {author} {\bibfnamefont {B.}~\bibnamefont {Pr{\'e}vost}}, \bibinfo
  {author} {\bibfnamefont {A.}~\bibnamefont {D{\'e}silets-Benoit}}, \bibinfo
  {author} {\bibfnamefont {H.~A.}\ \bibnamefont {Dabkowska}}, \bibinfo {author}
  {\bibfnamefont {O.}~\bibnamefont {Zaharko}}, \bibinfo {author} {\bibfnamefont
  {M.}~\bibnamefont {Frontzek}}, \bibinfo {author} {\bibfnamefont
  {R.}~\bibnamefont {Sibille}}, \bibinfo {author} {\bibfnamefont {A.~D.}\
  \bibnamefont {Bianchi}}, \ and\ \bibinfo {author} {\bibfnamefont
  {M.}~\bibnamefont {Kenzelmann}},\ }\href@noop {} {\bibfield  {journal}
  {\bibinfo  {journal} {Physical Review B}\ }\textbf {\bibinfo {volume} {95}},\
  \bibinfo {pages} {184436} (\bibinfo {year} {2017}{\natexlab{b}})}\BibitemShut
  {NoStop}%
\bibitem [{\citenamefont {Fennell}\ \emph
  {et~al.}(2014{\natexlab{b}})\citenamefont {Fennell}, \citenamefont
  {Kenzelmann}, \citenamefont {Roessli}, \citenamefont {Mutka}, \citenamefont
  {Ollivier}, \citenamefont {Ruminy}, \citenamefont {Stuhr}, \citenamefont
  {Zaharko}, \citenamefont {Bovo}, \citenamefont {Cervellino}, \citenamefont
  {Haas},\ and\ \citenamefont {Cava}}]{Fennell:2014gf}%
  \BibitemOpen
  \bibfield  {author} {\bibinfo {author} {\bibfnamefont {T.}~\bibnamefont
  {Fennell}}, \bibinfo {author} {\bibfnamefont {M.}~\bibnamefont {Kenzelmann}},
  \bibinfo {author} {\bibfnamefont {B.}~\bibnamefont {Roessli}}, \bibinfo
  {author} {\bibfnamefont {H.}~\bibnamefont {Mutka}}, \bibinfo {author}
  {\bibfnamefont {J.}~\bibnamefont {Ollivier}}, \bibinfo {author}
  {\bibfnamefont {M.}~\bibnamefont {Ruminy}}, \bibinfo {author} {\bibfnamefont
  {U.}~\bibnamefont {Stuhr}}, \bibinfo {author} {\bibfnamefont
  {O.}~\bibnamefont {Zaharko}}, \bibinfo {author} {\bibfnamefont
  {L.}~\bibnamefont {Bovo}}, \bibinfo {author} {\bibfnamefont {A.}~\bibnamefont
  {Cervellino}}, \bibinfo {author} {\bibfnamefont {M.~K.}\ \bibnamefont
  {Haas}}, \ and\ \bibinfo {author} {\bibfnamefont {R.~J.}\ \bibnamefont
  {Cava}},\ }\href@noop {} {\bibfield  {journal} {\bibinfo  {journal} {Physical
  Review Letters}\ }\textbf {\bibinfo {volume} {112}},\ \bibinfo {pages}
  {017203} (\bibinfo {year} {2014}{\natexlab{b}})}\BibitemShut {NoStop}%
\bibitem [{\citenamefont {Selke}(1988)}]{Selke:1988tx}%
  \BibitemOpen
  \bibfield  {author} {\bibinfo {author} {\bibfnamefont {W.}~\bibnamefont
  {Selke}},\ }\href@noop {} {\bibfield  {journal} {\bibinfo  {journal} {Physics
  Reports}\ }\textbf {\bibinfo {volume} {170}},\ \bibinfo {pages} {213}
  (\bibinfo {year} {1988})}\BibitemShut {NoStop}%
\bibitem [{\citenamefont {Mikeska}\ and\ \citenamefont
  {Kolezhuk}(2004)}]{Schollwock:2004wr}%
  \BibitemOpen
  \bibfield  {author} {\bibinfo {author} {\bibfnamefont {H.-J.}\ \bibnamefont
  {Mikeska}}\ and\ \bibinfo {author} {\bibfnamefont {A.~K.}\ \bibnamefont
  {Kolezhuk}},\ }\href@noop {} {\emph {\bibinfo {title} {{Quantum
  Magnetism}}}},\ edited by\ \bibinfo {editor} {\bibfnamefont {U.}~\bibnamefont
  {Schollw{\"o}ck}}, \bibinfo {editor} {\bibfnamefont {J.}~\bibnamefont
  {Richter}}, \bibinfo {editor} {\bibfnamefont {D.~J.~J.}\ \bibnamefont
  {Farnell}}, \ and\ \bibinfo {editor} {\bibfnamefont {R.~F.}\ \bibnamefont
  {Bishop}},\ \bibinfo {series} {Lecture Notes in Physics}, Vol.\ \bibinfo
  {volume} {645}\ (\bibinfo  {publisher} {Springer},\ \bibinfo {address}
  {Berlin Heidelberg},\ \bibinfo {year} {2004})\BibitemShut {NoStop}%
\bibitem [{\citenamefont {Ising}(1925)}]{Ising:1925eg}%
  \BibitemOpen
  \bibfield  {author} {\bibinfo {author} {\bibfnamefont {E.}~\bibnamefont
  {Ising}},\ }\href@noop {} {\bibfield  {journal} {\bibinfo  {journal}
  {Zeitschrift f{\"u}r Physik}\ }\textbf {\bibinfo {volume} {31}},\ \bibinfo
  {pages} {253} (\bibinfo {year} {1925})}\BibitemShut {NoStop}%
\bibitem [{\citenamefont {Heisenberg}(1928)}]{Heisenberg:1928jw}%
  \BibitemOpen
  \bibfield  {author} {\bibinfo {author} {\bibfnamefont {W.}~\bibnamefont
  {Heisenberg}},\ }\href@noop {} {\bibfield  {journal} {\bibinfo  {journal}
  {Zeitschrift f{\"u}r Physik}\ }\textbf {\bibinfo {volume} {49}},\ \bibinfo
  {pages} {619} (\bibinfo {year} {1928})}\BibitemShut {NoStop}%
\bibitem [{\citenamefont {Bethe}(1931)}]{Bethe:1931iz}%
  \BibitemOpen
  \bibfield  {author} {\bibinfo {author} {\bibfnamefont {H.}~\bibnamefont
  {Bethe}},\ }\href@noop {} {\bibfield  {journal} {\bibinfo  {journal}
  {Zeitschrift f{\"u}r Physik}\ }\textbf {\bibinfo {volume} {71}},\ \bibinfo
  {pages} {205} (\bibinfo {year} {1931})}\BibitemShut {NoStop}%
\bibitem [{\citenamefont {Mermin}\ and\ \citenamefont
  {Wagner}(1966)}]{Mermin:1966da}%
  \BibitemOpen
  \bibfield  {author} {\bibinfo {author} {\bibfnamefont {N.~D.}\ \bibnamefont
  {Mermin}}\ and\ \bibinfo {author} {\bibfnamefont {H.}~\bibnamefont
  {Wagner}},\ }\href@noop {} {\bibfield  {journal} {\bibinfo  {journal}
  {Physical Review Letters}\ }\textbf {\bibinfo {volume} {17}},\ \bibinfo
  {pages} {1133} (\bibinfo {year} {1966})}\BibitemShut {NoStop}%
\bibitem [{\citenamefont {Faddeev}\ and\ \citenamefont
  {Takhtajan}(1981)}]{Faddeev:1981cl}%
  \BibitemOpen
  \bibfield  {author} {\bibinfo {author} {\bibfnamefont {L.~D.}\ \bibnamefont
  {Faddeev}}\ and\ \bibinfo {author} {\bibfnamefont {L.~A.}\ \bibnamefont
  {Takhtajan}},\ }\href@noop {} {\bibfield  {journal} {\bibinfo  {journal}
  {Physics Letters A}\ }\textbf {\bibinfo {volume} {85}},\ \bibinfo {pages}
  {375} (\bibinfo {year} {1981})}\BibitemShut {NoStop}%
\bibitem [{\citenamefont {Haldane}(1983{\natexlab{a}})}]{Haldane:1983cl}%
  \BibitemOpen
  \bibfield  {author} {\bibinfo {author} {\bibfnamefont {F.~D.~M.}\
  \bibnamefont {Haldane}},\ }\href@noop {} {\bibfield  {journal} {\bibinfo
  {journal} {Physics Letters A}\ }\textbf {\bibinfo {volume} {93}},\ \bibinfo
  {pages} {464} (\bibinfo {year} {1983}{\natexlab{a}})}\BibitemShut {NoStop}%
\bibitem [{\citenamefont {Haldane}(1983{\natexlab{b}})}]{Haldane:1983ip}%
  \BibitemOpen
  \bibfield  {author} {\bibinfo {author} {\bibfnamefont {F.~D.~M.}\
  \bibnamefont {Haldane}},\ }\href@noop {} {\bibfield  {journal} {\bibinfo
  {journal} {Physical Review Letters}\ }\textbf {\bibinfo {volume} {50}},\
  \bibinfo {pages} {1153} (\bibinfo {year} {1983}{\natexlab{b}})}\BibitemShut
  {NoStop}%
\bibitem [{\citenamefont {Dagotto}\ and\ \citenamefont
  {Rice}(1996)}]{Dagotto:1996jl}%
  \BibitemOpen
  \bibfield  {author} {\bibinfo {author} {\bibfnamefont {E.}~\bibnamefont
  {Dagotto}}\ and\ \bibinfo {author} {\bibfnamefont {T.~M.}\ \bibnamefont
  {Rice}},\ }\href@noop {} {\bibfield  {journal} {\bibinfo  {journal}
  {Science}\ }\textbf {\bibinfo {volume} {271}},\ \bibinfo {pages} {618}
  (\bibinfo {year} {1996})}\BibitemShut {NoStop}%
\bibitem [{\citenamefont {Affleck}(1990)}]{Affleck:1990cw}%
  \BibitemOpen
  \bibfield  {author} {\bibinfo {author} {\bibfnamefont {I.}~\bibnamefont
  {Affleck}},\ }\href@noop {} {\bibfield  {journal} {\bibinfo  {journal}
  {Physical Review B}\ }\textbf {\bibinfo {volume} {41}},\ \bibinfo {pages}
  {6697} (\bibinfo {year} {1990})}\BibitemShut {NoStop}%
\bibitem [{\citenamefont {Oshikawa}\ \emph {et~al.}(1997)\citenamefont
  {Oshikawa}, \citenamefont {Yamanaka},\ and\ \citenamefont
  {Affleck}}]{Oshikawa:1997fx}%
  \BibitemOpen
  \bibfield  {author} {\bibinfo {author} {\bibfnamefont {M.}~\bibnamefont
  {Oshikawa}}, \bibinfo {author} {\bibfnamefont {M.}~\bibnamefont {Yamanaka}},
  \ and\ \bibinfo {author} {\bibfnamefont {I.}~\bibnamefont {Affleck}},\
  }\href@noop {} {\bibfield  {journal} {\bibinfo  {journal} {Physical Review
  Letters}\ }\textbf {\bibinfo {volume} {78}},\ \bibinfo {pages} {1984}
  (\bibinfo {year} {1997})}\BibitemShut {NoStop}%
\bibitem [{\citenamefont {Heidrich-Meisner}\ \emph {et~al.}(2007)\citenamefont
  {Heidrich-Meisner}, \citenamefont {Sergienko}, \citenamefont {Feiguin},\ and\
  \citenamefont {Dagotto}}]{HeidrichMeisner:2007ey}%
  \BibitemOpen
  \bibfield  {author} {\bibinfo {author} {\bibfnamefont {F.}~\bibnamefont
  {Heidrich-Meisner}}, \bibinfo {author} {\bibfnamefont {I.~A.}\ \bibnamefont
  {Sergienko}}, \bibinfo {author} {\bibfnamefont {A.~E.}\ \bibnamefont
  {Feiguin}}, \ and\ \bibinfo {author} {\bibfnamefont {E.~R.}\ \bibnamefont
  {Dagotto}},\ }\href@noop {} {\bibfield  {journal} {\bibinfo  {journal}
  {Physical Review B}\ }\textbf {\bibinfo {volume} {75}},\ \bibinfo {pages}
  {064413} (\bibinfo {year} {2007})}\BibitemShut {NoStop}%
\bibitem [{\citenamefont {McCulloch}\ \emph {et~al.}(2008)\citenamefont
  {McCulloch}, \citenamefont {Kube}, \citenamefont {Kurz}, \citenamefont
  {Kleine}, \citenamefont {Schollw{\"o}ck},\ and\ \citenamefont
  {Kolezhuk}}]{McCulloch:2008cw}%
  \BibitemOpen
  \bibfield  {author} {\bibinfo {author} {\bibfnamefont {I.~P.}\ \bibnamefont
  {McCulloch}}, \bibinfo {author} {\bibfnamefont {R.}~\bibnamefont {Kube}},
  \bibinfo {author} {\bibfnamefont {M.}~\bibnamefont {Kurz}}, \bibinfo {author}
  {\bibfnamefont {A.}~\bibnamefont {Kleine}}, \bibinfo {author} {\bibfnamefont
  {U.}~\bibnamefont {Schollw{\"o}ck}}, \ and\ \bibinfo {author} {\bibfnamefont
  {A.~K.}\ \bibnamefont {Kolezhuk}},\ }\href@noop {} {\bibfield  {journal}
  {\bibinfo  {journal} {Physical Review B}\ }\textbf {\bibinfo {volume} {77}},\
  \bibinfo {pages} {094404} (\bibinfo {year} {2008})}\BibitemShut {NoStop}%
\bibitem [{\citenamefont {Okunishi}\ and\ \citenamefont
  {Tonegawa}(2003)}]{Okunishi:2003ux}%
  \BibitemOpen
  \bibfield  {author} {\bibinfo {author} {\bibfnamefont {K.}~\bibnamefont
  {Okunishi}}\ and\ \bibinfo {author} {\bibfnamefont {T.}~\bibnamefont
  {Tonegawa}},\ }\href@noop {} {\bibfield  {journal} {\bibinfo  {journal}
  {Physical Review B}\ }\textbf {\bibinfo {volume} {68}},\ \bibinfo {pages}
  {224422} (\bibinfo {year} {2003})}\BibitemShut {NoStop}%
\bibitem [{\citenamefont {Doi}\ \emph {et~al.}(2006)\citenamefont {Doi},
  \citenamefont {Nakamori},\ and\ \citenamefont {Hinatsu}}]{Doi:2006ue}%
  \BibitemOpen
  \bibfield  {author} {\bibinfo {author} {\bibfnamefont {Y.}~\bibnamefont
  {Doi}}, \bibinfo {author} {\bibfnamefont {W.}~\bibnamefont {Nakamori}}, \
  and\ \bibinfo {author} {\bibfnamefont {Y.}~\bibnamefont {Hinatsu}},\
  }\href@noop {} {\bibfield  {journal} {\bibinfo  {journal} {Journal of
  Physics: Condensed Matter}\ }\textbf {\bibinfo {volume} {18}},\ \bibinfo
  {pages} {333} (\bibinfo {year} {2006})}\BibitemShut {NoStop}%
\bibitem [{\citenamefont {Besara}\ \emph {et~al.}(2014)\citenamefont {Besara},
  \citenamefont {Lundberg}, \citenamefont {Sun}, \citenamefont {Ramirez},
  \citenamefont {Dong}, \citenamefont {Whalen}, \citenamefont {Vasquez},
  \citenamefont {Herrera}, \citenamefont {Allen}, \citenamefont {Davidson},\
  and\ \citenamefont {Siegrist}}]{Besara:2014kx}%
  \BibitemOpen
  \bibfield  {author} {\bibinfo {author} {\bibfnamefont {T.}~\bibnamefont
  {Besara}}, \bibinfo {author} {\bibfnamefont {M.~S.}\ \bibnamefont
  {Lundberg}}, \bibinfo {author} {\bibfnamefont {J.}~\bibnamefont {Sun}},
  \bibinfo {author} {\bibfnamefont {D.}~\bibnamefont {Ramirez}}, \bibinfo
  {author} {\bibfnamefont {L.}~\bibnamefont {Dong}}, \bibinfo {author}
  {\bibfnamefont {J.~B.}\ \bibnamefont {Whalen}}, \bibinfo {author}
  {\bibfnamefont {R.}~\bibnamefont {Vasquez}}, \bibinfo {author} {\bibfnamefont
  {F.}~\bibnamefont {Herrera}}, \bibinfo {author} {\bibfnamefont {J.~R.}\
  \bibnamefont {Allen}}, \bibinfo {author} {\bibfnamefont {M.~W.}\ \bibnamefont
  {Davidson}}, \ and\ \bibinfo {author} {\bibfnamefont {T.}~\bibnamefont
  {Siegrist}},\ }\href@noop {} {\bibfield  {journal} {\bibinfo  {journal}
  {Progress in Solid State Chemistry}\ }\textbf {\bibinfo {volume} {42}},\
  \bibinfo {pages} {23} (\bibinfo {year} {2014})}\BibitemShut {NoStop}%
\bibitem [{\citenamefont {Bewley}\ \emph {et~al.}(2006)\citenamefont {Bewley},
  \citenamefont {Eccleston}, \citenamefont {McEwen}, \citenamefont {Hayden},
  \citenamefont {Dove}, \citenamefont {Bennington}, \citenamefont {Treadgold},\
  and\ \citenamefont {Coleman}}]{Bewley:2006bl}%
  \BibitemOpen
  \bibfield  {author} {\bibinfo {author} {\bibfnamefont {R.~I.}\ \bibnamefont
  {Bewley}}, \bibinfo {author} {\bibfnamefont {R.~S.}\ \bibnamefont
  {Eccleston}}, \bibinfo {author} {\bibfnamefont {K.~A.}\ \bibnamefont
  {McEwen}}, \bibinfo {author} {\bibfnamefont {S.~M.}\ \bibnamefont {Hayden}},
  \bibinfo {author} {\bibfnamefont {M.~T.}\ \bibnamefont {Dove}}, \bibinfo
  {author} {\bibfnamefont {S.~M.}\ \bibnamefont {Bennington}}, \bibinfo
  {author} {\bibfnamefont {J.~R.}\ \bibnamefont {Treadgold}}, \ and\ \bibinfo
  {author} {\bibfnamefont {R.~L.~S.}\ \bibnamefont {Coleman}},\ }\href@noop {}
  {\bibfield  {journal} {\bibinfo  {journal} {Physica B: Condensed Matter}\
  }\textbf {\bibinfo {volume} {385-386}},\ \bibinfo {pages} {1029} (\bibinfo
  {year} {2006})}\BibitemShut {NoStop}%
\bibitem [{\citenamefont {Schmitt}\ and\ \citenamefont
  {Ouladdiaf}(1998)}]{Schmitt:1998by}%
  \BibitemOpen
  \bibfield  {author} {\bibinfo {author} {\bibfnamefont {D.}~\bibnamefont
  {Schmitt}}\ and\ \bibinfo {author} {\bibfnamefont {B.}~\bibnamefont
  {Ouladdiaf}},\ }\href@noop {} {\bibfield  {journal} {\bibinfo  {journal}
  {Journal of Applied Crystallography}\ }\textbf {\bibinfo {volume} {31}},\
  \bibinfo {pages} {620} (\bibinfo {year} {1998})}\BibitemShut {NoStop}%
\bibitem [{\citenamefont {Uldry}\ \emph {et~al.}(2012)\citenamefont {Uldry},
  \citenamefont {Vernay},\ and\ \citenamefont {Delley}}]{Uldry:2012ky}%
  \BibitemOpen
  \bibfield  {author} {\bibinfo {author} {\bibfnamefont {A.}~\bibnamefont
  {Uldry}}, \bibinfo {author} {\bibfnamefont {F.}~\bibnamefont {Vernay}}, \
  and\ \bibinfo {author} {\bibfnamefont {B.}~\bibnamefont {Delley}},\
  }\href@noop {} {\bibfield  {journal} {\bibinfo  {journal} {Physical Review
  B}\ }\textbf {\bibinfo {volume} {85}},\ \bibinfo {pages} {125133} (\bibinfo
  {year} {2012})}\BibitemShut {NoStop}%
\bibitem [{\citenamefont {Fischer}\ \emph {et~al.}(2000)\citenamefont
  {Fischer}, \citenamefont {Frey}, \citenamefont {Koch}, \citenamefont
  {K{\"o}nnecke}, \citenamefont {Pomjakushin}, \citenamefont {Schefer},
  \citenamefont {Thut}, \citenamefont {Schlumpf}, \citenamefont {B{\"u}rge},
  \citenamefont {Greuter}, \citenamefont {Bondt},\ and\ \citenamefont
  {Berruyer}}]{Fischer:2000ja}%
  \BibitemOpen
  \bibfield  {author} {\bibinfo {author} {\bibfnamefont {P.}~\bibnamefont
  {Fischer}}, \bibinfo {author} {\bibfnamefont {G.}~\bibnamefont {Frey}},
  \bibinfo {author} {\bibfnamefont {M.}~\bibnamefont {Koch}}, \bibinfo {author}
  {\bibfnamefont {M.}~\bibnamefont {K{\"o}nnecke}}, \bibinfo {author}
  {\bibfnamefont {V.}~\bibnamefont {Pomjakushin}}, \bibinfo {author}
  {\bibfnamefont {J.}~\bibnamefont {Schefer}}, \bibinfo {author} {\bibfnamefont
  {R.}~\bibnamefont {Thut}}, \bibinfo {author} {\bibfnamefont {N.}~\bibnamefont
  {Schlumpf}}, \bibinfo {author} {\bibfnamefont {R.}~\bibnamefont {B{\"u}rge}},
  \bibinfo {author} {\bibfnamefont {U.}~\bibnamefont {Greuter}}, \bibinfo
  {author} {\bibfnamefont {S.}~\bibnamefont {Bondt}}, \ and\ \bibinfo {author}
  {\bibfnamefont {E.}~\bibnamefont {Berruyer}},\ }\href@noop {} {\bibfield
  {journal} {\bibinfo  {journal} {Physica B: Condensed Matter}\ }\textbf
  {\bibinfo {volume} {276-278}},\ \bibinfo {pages} {146} (\bibinfo {year}
  {2000})}\BibitemShut {NoStop}%
\bibitem [{\citenamefont
  {Rodr{\'\i}guez-Carvajal}(1993)}]{RodriguezCarvajal:1993cf}%
  \BibitemOpen
  \bibfield  {author} {\bibinfo {author} {\bibfnamefont {J.}~\bibnamefont
  {Rodr{\'\i}guez-Carvajal}},\ }\href@noop {} {\bibfield  {journal} {\bibinfo
  {journal} {Physica B: Condensed Matter}\ }\textbf {\bibinfo {volume} {192}},\
  \bibinfo {pages} {55} (\bibinfo {year} {1993})}\BibitemShut {NoStop}%
\bibitem [{\citenamefont {Marder}(2010)}]{Marder:2010ta}%
  \BibitemOpen
  \bibfield  {author} {\bibinfo {author} {\bibfnamefont {M.~P.}\ \bibnamefont
  {Marder}},\ }\href@noop {} {\emph {\bibinfo {title} {{Condensed Matter
  Physics}}}}\ (\bibinfo  {publisher} {John Wiley {\&} Sons},\ \bibinfo {year}
  {2010})\BibitemShut {NoStop}%
\bibitem [{\citenamefont {Li}\ \emph {et~al.}(2014)\citenamefont {Li},
  \citenamefont {Zhang}, \citenamefont {Senyshyn}, \citenamefont {Wildes},
  \citenamefont {Schmalzl}, \citenamefont {Schmidt}, \citenamefont {Boehm},
  \citenamefont {Ressouche}, \citenamefont {Hou}, \citenamefont {Meuffels},
  \citenamefont {Roth},\ and\ \citenamefont {Br{\"u}ckel}}]{Li:2014gs}%
  \BibitemOpen
  \bibfield  {author} {\bibinfo {author} {\bibfnamefont {H.-F.}\ \bibnamefont
  {Li}}, \bibinfo {author} {\bibfnamefont {C.}~\bibnamefont {Zhang}}, \bibinfo
  {author} {\bibfnamefont {A.}~\bibnamefont {Senyshyn}}, \bibinfo {author}
  {\bibfnamefont {A.}~\bibnamefont {Wildes}}, \bibinfo {author} {\bibfnamefont
  {K.}~\bibnamefont {Schmalzl}}, \bibinfo {author} {\bibfnamefont
  {W.}~\bibnamefont {Schmidt}}, \bibinfo {author} {\bibfnamefont
  {M.}~\bibnamefont {Boehm}}, \bibinfo {author} {\bibfnamefont
  {E.}~\bibnamefont {Ressouche}}, \bibinfo {author} {\bibfnamefont
  {B.}~\bibnamefont {Hou}}, \bibinfo {author} {\bibfnamefont {P.}~\bibnamefont
  {Meuffels}}, \bibinfo {author} {\bibfnamefont {G.}~\bibnamefont {Roth}}, \
  and\ \bibinfo {author} {\bibfnamefont {T.}~\bibnamefont {Br{\"u}ckel}},\
  }\href@noop {} {\bibfield  {journal} {\bibinfo  {journal} {Frontiers in
  Physics}\ }\textbf {\bibinfo {volume} {2}},\ \bibinfo {pages} {42} (\bibinfo
  {year} {2014})}\BibitemShut {NoStop}%
\bibitem [{\citenamefont {Malkin}\ \emph {et~al.}(2015)\citenamefont {Malkin},
  \citenamefont {Nikitin}, \citenamefont {Mumdzhi}, \citenamefont {Zverev},
  \citenamefont {Yusupov}, \citenamefont {Gilmutdinov}, \citenamefont
  {Batulin}, \citenamefont {Gabbasov}, \citenamefont {Kiiamov}, \citenamefont
  {Adroja}, \citenamefont {Young},\ and\ \citenamefont
  {Petrenko}}]{Malkin:2015eh}%
  \BibitemOpen
  \bibfield  {author} {\bibinfo {author} {\bibfnamefont {B.~Z.}\ \bibnamefont
  {Malkin}}, \bibinfo {author} {\bibfnamefont {S.~I.}\ \bibnamefont {Nikitin}},
  \bibinfo {author} {\bibfnamefont {I.~E.}\ \bibnamefont {Mumdzhi}}, \bibinfo
  {author} {\bibfnamefont {D.~G.}\ \bibnamefont {Zverev}}, \bibinfo {author}
  {\bibfnamefont {R.~V.}\ \bibnamefont {Yusupov}}, \bibinfo {author}
  {\bibfnamefont {I.~F.}\ \bibnamefont {Gilmutdinov}}, \bibinfo {author}
  {\bibfnamefont {R.}~\bibnamefont {Batulin}}, \bibinfo {author} {\bibfnamefont
  {B.~F.}\ \bibnamefont {Gabbasov}}, \bibinfo {author} {\bibfnamefont {A.~G.}\
  \bibnamefont {Kiiamov}}, \bibinfo {author} {\bibfnamefont {D.~T.}\
  \bibnamefont {Adroja}}, \bibinfo {author} {\bibfnamefont {O.}~\bibnamefont
  {Young}}, \ and\ \bibinfo {author} {\bibfnamefont {O.~A.}\ \bibnamefont
  {Petrenko}},\ }\href@noop {} {\bibfield  {journal} {\bibinfo  {journal}
  {Physical Review B}\ }\textbf {\bibinfo {volume} {92}},\ \bibinfo {pages}
  {094415} (\bibinfo {year} {2015})}\BibitemShut {NoStop}%
\bibitem [{\citenamefont {Warren}(1941)}]{Warren:1941cj}%
  \BibitemOpen
  \bibfield  {author} {\bibinfo {author} {\bibfnamefont {B.~E.}\ \bibnamefont
  {Warren}},\ }\href@noop {} {\bibfield  {journal} {\bibinfo  {journal}
  {Physical Review}\ }\textbf {\bibinfo {volume} {59}},\ \bibinfo {pages} {693}
  (\bibinfo {year} {1941})}\BibitemShut {NoStop}%
\bibitem [{\citenamefont {Jones}(1949)}]{Jones:1949cx}%
  \BibitemOpen
  \bibfield  {author} {\bibinfo {author} {\bibfnamefont {R.~C.}\ \bibnamefont
  {Jones}},\ }\href@noop {} {\bibfield  {journal} {\bibinfo  {journal} {Acta
  Crystallographica}\ }\textbf {\bibinfo {volume} {2}},\ \bibinfo {pages} {252}
  (\bibinfo {year} {1949})}\BibitemShut {NoStop}%
\bibitem [{\citenamefont {Yamani}\ \emph {et~al.}(2010)\citenamefont {Yamani},
  \citenamefont {Tun},\ and\ \citenamefont {Ryan}}]{Yamani:2010vg}%
  \BibitemOpen
  \bibfield  {author} {\bibinfo {author} {\bibfnamefont {Z.}~\bibnamefont
  {Yamani}}, \bibinfo {author} {\bibfnamefont {Z.}~\bibnamefont {Tun}}, \ and\
  \bibinfo {author} {\bibfnamefont {D.~H.}\ \bibnamefont {Ryan}},\ }\href@noop
  {} {\bibfield  {journal} {\bibinfo  {journal} {Canadian Journal of Physics}\
  }\textbf {\bibinfo {volume} {88}},\ \bibinfo {pages} {771} (\bibinfo {year}
  {2010})}\BibitemShut {NoStop}%
\bibitem [{\citenamefont {Stephenson}(1970)}]{Stephenson:1970cs}%
  \BibitemOpen
  \bibfield  {author} {\bibinfo {author} {\bibfnamefont {J.}~\bibnamefont
  {Stephenson}},\ }\href@noop {} {\bibfield  {journal} {\bibinfo  {journal}
  {Canadian Journal of Physics}\ }\textbf {\bibinfo {volume} {48}},\ \bibinfo
  {pages} {1724} (\bibinfo {year} {1970})}\BibitemShut {NoStop}%
\bibitem [{\citenamefont {Haraldsen}\ \emph {et~al.}(2005)\citenamefont
  {Haraldsen}, \citenamefont {Barnes},\ and\ \citenamefont
  {Musfeldt}}]{Haraldsen:2005kt}%
  \BibitemOpen
  \bibfield  {author} {\bibinfo {author} {\bibfnamefont {J.~T.}\ \bibnamefont
  {Haraldsen}}, \bibinfo {author} {\bibfnamefont {T.}~\bibnamefont {Barnes}}, \
  and\ \bibinfo {author} {\bibfnamefont {J.~L.}\ \bibnamefont {Musfeldt}},\
  }\href@noop {} {\bibfield  {journal} {\bibinfo  {journal} {Physical Review
  B}\ }\textbf {\bibinfo {volume} {71}},\ \bibinfo {pages} {064403} (\bibinfo
  {year} {2005})}\BibitemShut {NoStop}%
\bibitem [{\citenamefont {Aczel}\ \emph {et~al.}(2014)\citenamefont {Aczel},
  \citenamefont {Li}, \citenamefont {Garlea}, \citenamefont {Yan},
  \citenamefont {Weickert}, \citenamefont {Jaime}, \citenamefont {Maiorov},
  \citenamefont {Movshovich}, \citenamefont {Civale}, \citenamefont {Keppens},\
  and\ \citenamefont {Mandrus}}]{Aczel:2014ko}%
  \BibitemOpen
  \bibfield  {author} {\bibinfo {author} {\bibfnamefont {A.~A.}\ \bibnamefont
  {Aczel}}, \bibinfo {author} {\bibfnamefont {L.}~\bibnamefont {Li}}, \bibinfo
  {author} {\bibfnamefont {V.~O.}\ \bibnamefont {Garlea}}, \bibinfo {author}
  {\bibfnamefont {J.~Q.}\ \bibnamefont {Yan}}, \bibinfo {author} {\bibfnamefont
  {F.}~\bibnamefont {Weickert}}, \bibinfo {author} {\bibfnamefont
  {M.}~\bibnamefont {Jaime}}, \bibinfo {author} {\bibfnamefont
  {B.}~\bibnamefont {Maiorov}}, \bibinfo {author} {\bibfnamefont
  {R.}~\bibnamefont {Movshovich}}, \bibinfo {author} {\bibfnamefont
  {L.}~\bibnamefont {Civale}}, \bibinfo {author} {\bibfnamefont
  {V.}~\bibnamefont {Keppens}}, \ and\ \bibinfo {author} {\bibfnamefont
  {D.}~\bibnamefont {Mandrus}},\ }\href@noop {} {\bibfield  {journal} {\bibinfo
   {journal} {Physical Review B}\ }\textbf {\bibinfo {volume} {90}},\ \bibinfo
  {pages} {134403} (\bibinfo {year} {2014})}\BibitemShut {NoStop}%
\bibitem [{\citenamefont {Aczel}\ \emph {et~al.}(2015)\citenamefont {Aczel},
  \citenamefont {Li}, \citenamefont {Garlea}, \citenamefont {Yan},
  \citenamefont {Weickert}, \citenamefont {Zapf}, \citenamefont {Movshovich},
  \citenamefont {Jaime}, \citenamefont {Baker}, \citenamefont {Keppens},\ and\
  \citenamefont {Mandrus}}]{Aczel:2015ko}%
  \BibitemOpen
  \bibfield  {author} {\bibinfo {author} {\bibfnamefont {A.~A.}\ \bibnamefont
  {Aczel}}, \bibinfo {author} {\bibfnamefont {L.}~\bibnamefont {Li}}, \bibinfo
  {author} {\bibfnamefont {V.~O.}\ \bibnamefont {Garlea}}, \bibinfo {author}
  {\bibfnamefont {J.~Q.}\ \bibnamefont {Yan}}, \bibinfo {author} {\bibfnamefont
  {F.}~\bibnamefont {Weickert}}, \bibinfo {author} {\bibfnamefont {V.~S.}\
  \bibnamefont {Zapf}}, \bibinfo {author} {\bibfnamefont {R.}~\bibnamefont
  {Movshovich}}, \bibinfo {author} {\bibfnamefont {M.}~\bibnamefont {Jaime}},
  \bibinfo {author} {\bibfnamefont {P.~J.}\ \bibnamefont {Baker}}, \bibinfo
  {author} {\bibfnamefont {V.}~\bibnamefont {Keppens}}, \ and\ \bibinfo
  {author} {\bibfnamefont {D.}~\bibnamefont {Mandrus}},\ }\href@noop {}
  {\bibfield  {journal} {\bibinfo  {journal} {Physical Review B}\ }\textbf
  {\bibinfo {volume} {92}},\ \bibinfo {pages} {041110} (\bibinfo {year}
  {2015})}\BibitemShut {NoStop}%
\bibitem [{\citenamefont {Young}\ \emph {et~al.}(2014)\citenamefont {Young},
  \citenamefont {Balakrishnan}, \citenamefont {Lees},\ and\ \citenamefont
  {Petrenko}}]{Young:2014bc}%
  \BibitemOpen
  \bibfield  {author} {\bibinfo {author} {\bibfnamefont {O.}~\bibnamefont
  {Young}}, \bibinfo {author} {\bibfnamefont {G.}~\bibnamefont {Balakrishnan}},
  \bibinfo {author} {\bibfnamefont {M.~R.}\ \bibnamefont {Lees}}, \ and\
  \bibinfo {author} {\bibfnamefont {O.~A.}\ \bibnamefont {Petrenko}},\
  }\href@noop {} {\bibfield  {journal} {\bibinfo  {journal} {Physical Review
  B}\ }\textbf {\bibinfo {volume} {90}},\ \bibinfo {pages} {094421} (\bibinfo
  {year} {2014})}\BibitemShut {NoStop}%
\bibitem [{\citenamefont {Brooks-Bartlett}\ \emph {et~al.}(2014)\citenamefont
  {Brooks-Bartlett}, \citenamefont {Banks}, \citenamefont {Jaubert},
  \citenamefont {Harman-Clarke},\ and\ \citenamefont
  {Holdsworth}}]{BrooksBartlett:2014kf}%
  \BibitemOpen
  \bibfield  {author} {\bibinfo {author} {\bibfnamefont {M.~E.}\ \bibnamefont
  {Brooks-Bartlett}}, \bibinfo {author} {\bibfnamefont {S.~T.}\ \bibnamefont
  {Banks}}, \bibinfo {author} {\bibfnamefont {L.~D.~C.}\ \bibnamefont
  {Jaubert}}, \bibinfo {author} {\bibfnamefont {A.}~\bibnamefont
  {Harman-Clarke}}, \ and\ \bibinfo {author} {\bibfnamefont {P.~C.~W.}\
  \bibnamefont {Holdsworth}},\ }\href@noop {} {\bibfield  {journal} {\bibinfo
  {journal} {Physical Review X}\ }\textbf {\bibinfo {volume} {4}},\ \bibinfo
  {pages} {011007} (\bibinfo {year} {2014})}\BibitemShut {NoStop}%
\bibitem [{\citenamefont {Petit}\ \emph {et~al.}(2016)\citenamefont {Petit},
  \citenamefont {Lhotel}, \citenamefont {Canals}, \citenamefont
  {Ciomaga~Hatnean}, \citenamefont {Ollivier}, \citenamefont {Mutka},
  \citenamefont {Ressouche}, \citenamefont {Wildes}, \citenamefont {Lees},\
  and\ \citenamefont {Balakrishnan}}]{Petit:2016dg}%
  \BibitemOpen
  \bibfield  {author} {\bibinfo {author} {\bibfnamefont {S.}~\bibnamefont
  {Petit}}, \bibinfo {author} {\bibfnamefont {E.}~\bibnamefont {Lhotel}},
  \bibinfo {author} {\bibfnamefont {B.}~\bibnamefont {Canals}}, \bibinfo
  {author} {\bibfnamefont {M.}~\bibnamefont {Ciomaga~Hatnean}}, \bibinfo
  {author} {\bibfnamefont {J.}~\bibnamefont {Ollivier}}, \bibinfo {author}
  {\bibfnamefont {H.}~\bibnamefont {Mutka}}, \bibinfo {author} {\bibfnamefont
  {E.}~\bibnamefont {Ressouche}}, \bibinfo {author} {\bibfnamefont {A.~R.}\
  \bibnamefont {Wildes}}, \bibinfo {author} {\bibfnamefont {M.~R.}\
  \bibnamefont {Lees}}, \ and\ \bibinfo {author} {\bibfnamefont
  {G.}~\bibnamefont {Balakrishnan}},\ }\href@noop {} {\bibfield  {journal}
  {\bibinfo  {journal} {Nature Physics}\ }\textbf {\bibinfo {volume} {12}},\
  \bibinfo {pages} {746} (\bibinfo {year} {2016})}\BibitemShut {NoStop}%
\bibitem [{\citenamefont {Gauthier}\ \emph
  {et~al.}(2017{\natexlab{c}})\citenamefont {Gauthier}, \citenamefont
  {Pr{\'e}vost}, \citenamefont {Amato}, \citenamefont {Baines}, \citenamefont
  {Pomjakushin}, \citenamefont {Bianchi}, \citenamefont {Cava},\ and\
  \citenamefont {Kenzelmann}}]{Gauthier:2017uh}%
  \BibitemOpen
  \bibfield  {author} {\bibinfo {author} {\bibfnamefont {N.}~\bibnamefont
  {Gauthier}}, \bibinfo {author} {\bibfnamefont {B.}~\bibnamefont
  {Pr{\'e}vost}}, \bibinfo {author} {\bibfnamefont {A.}~\bibnamefont {Amato}},
  \bibinfo {author} {\bibfnamefont {C.}~\bibnamefont {Baines}}, \bibinfo
  {author} {\bibfnamefont {V.}~\bibnamefont {Pomjakushin}}, \bibinfo {author}
  {\bibfnamefont {A.~D.}\ \bibnamefont {Bianchi}}, \bibinfo {author}
  {\bibfnamefont {R.~J.}\ \bibnamefont {Cava}}, \ and\ \bibinfo {author}
  {\bibfnamefont {M.}~\bibnamefont {Kenzelmann}},\ }\href@noop {} {\bibfield
  {journal} {\bibinfo  {journal} {Journal of Physics: Conference Series}\
  }\textbf {\bibinfo {volume} {828}},\ \bibinfo {pages} {012014} (\bibinfo
  {year} {2017}{\natexlab{c}})}\BibitemShut {NoStop}%
\end{thebibliography}%

\end{document}